\documentclass[10pt,final,twocolumn,a4paper,twoside]{IEEEtran}\newcommand{\duecol}{1}

\usepackage{ifthen}
\usepackage{amsmath}
\usepackage{theorem}
\usepackage{xspace}
\usepackage{subfigure}
\usepackage{calc}
\usepackage{amsfonts}
\usepackage{arydshln}

\newcommand{\unwmfperriga}[4]
{
\ifthenelse{\duecol=0}{\begin{figure}[htb!]}{\begin{figure}[htb!]}
\centerwmf{#1}{#2}{#3} #4
\ifthenelse{\duecol=0}{\end{figure}}{\end{figure}}
}

\newcommand{\duewmfperriga}[6]
{
\ifthenelse{\duecol=0}{\begin{figure}[htb!]}{\begin{figure*}[htb!]}
\vspace*{#2}\relax\noindent%
\ifthenelse{\duecol=0}{\hspace*{\duecolhindent}}{}%
\protect\special{wmf:#4 x=#1 y=#2}\hspace*{#3}%
\protect\special{wmf:#5 x=#1 y=#2}%
\vspace*{-1.em}{#6}
\ifthenelse{\duecol=1}{\vspace*{-1.em}}{\vspace*{-1.75em}}
\ifthenelse{\duecol=0}{\end{figure}}{\end{figure*}}
}

\newcommand{\artworks}{0}
\newcommand{\figures}{1}
\usepackage{ifthen}

\newcommand{\trytoturnpage}{\vspace*{20em}\par\noindent}
\newcommand{\myfig}[1]{\ifthenelse{\artworks=1}{\begin{figure}[f]\trytoturnpage}{\begin{figure}[#1]}}
\newcommand{\mytab}[1]{\ifthenelse{\artworks=1}{\begin{table}[f]\trytoturnpage}{\begin{table}[#1]}}
\newcommand{\myfigstar}[1]{\ifthenelse{\artworks=1}{\begin{figure*}[f]\trytoturnpage}{\begin{figure*}[#1]}}
\newcommand{\mytabstar}[1]{\ifthenelse{\artworks=1}{\begin{table*}[f]\trytoturnpage}{\begin{table*}[#1]}}

\newcommand{\mycaption}[1]{\ifthenelse{\artworks=1}{\vspace*{10em}\caption{#1}}{\caption{#1}}}

\newcommand{\myfigend}{\ifthenelse{\artworks=1}{\trytoturnpage\end{figure}}{\end{figure}}}
\newcommand{\myfigstarend}{\ifthenelse{\artworks=1}{\trytoturnpage\end{figure*}}{\end{figure*}}}

\newcommand{\mytabend}{\ifthenelse{\artworks=1}{\trytoturnpage\end{table}}{\end{table}}}
\newcommand{\mytabstarend}{\ifthenelse{\artworks=1}{\trytoturnpage\end{table*}}{\end{table*}}}

\newcommand{\mycenterwmf}[3]{\ifthenelse{\figures=1}{\centerwmf{#1}{#2}{#3}}{\vskip#2\medskip}}
\newcommand{\myspecial}[1]{\ifthenelse{\figures=1}{\special{#1}}{}}
\newcommand{\mycentereps}[3]{\ifthenelse{\figures=1}{\centereps{#1}{#2}{#3}}{\vskip#2\medskip}}

\newcommand{\ttt}{\ensuremath^{\scriptscriptstyle \mathrm{T}}}

\newcommand{\tras}{\ensuremath^{\scriptscriptstyle \mathrm{T}}}
\newcommand{\vet}[1]{{\rm \bf #1}}

\newcommand{\E}[1]{{\rm E}\left\{#1\right\}}

\newcommand{\ltextit}[1]{{\textit{#1}}}

\newcommand{\lmathcal}[1]{{\mathcal{#1}}}

\newcommand{\MSE}{\ensuremath{\text{MSE}}\xspace}

\newcommand{\ggdef}{\mathop {=} \limits^{\text{def}}}

\DeclareMathOperator{\Var}{Var}

{
\theoremstyle{break}
\theoremheaderfont{\normalfont\scshape}

}

\newcommand{\wrt}{w.r.t.\@\xspace}

\newcommand{\ie}{{\textit{i.e.}}\@\xspace}

\newcommand{\eg}{{\textit{e.g.}}\@\xspace}

\newcommand{\gVar}[1]{\Var\left\{#1\right\}}

\newsavebox{\fminibox}
\newlength{\fminilength}

\newcommand{\vety}{\vet{y}}

\newcommand{\unbmpperriga}[4]
{
\ifthenelse{\duecol=0}{\myfig{htb!}}{\myfig{htb!}}
\centerbmp{#1}{#2}{#3} #4
\ifthenelse{\duecol=0}{\myfigend}{\myfigend}
}

\newcommand{\unepsperriga}[4]
{
\ifthenelse{\duecol=0}{\myfig{htb!}}{\myfig{htb!}}
\mycentereps{#1}{#2}{#3}
#4\ifthenelse{\duecol=1}{\vspace*{-0.em}}{}
\ifthenelse{\duecol=0}{\myfigend}{\myfigend}
}

\newcommand{\duebmpperriga}[6]
{
\ifthenelse{\duecol=0}{\myfig{htb!}}{\myfig{htb!}}
\vspace*{#2}\relax\noindent
\ifthenelse{\duecol=0}{\hspace*{\duecolhindent}}{}%
\protect\myspecial{bmp:#4 x=#1 y=#2}\hspace*{#3}%
\protect\myspecial{bmp:#5 x=#1 y=#2}%
{#6}
\ifthenelse{\duecol=0}{\myfigend}{\myfigend}
}

\newcommand{\trebmpperriga}[7]
{
\ifthenelse{\duecol=0}{\myfig{htb!}}{\myfigstar{htb!}}
\vspace*{#2}\relax\noindent
\ifthenelse{\duecol=0}{\hspace*{\duecolhindent}}{}%
\protect\myspecial{bmp:#4 x=#1 y=#2}\hspace*{#3}%
\protect\myspecial{bmp:#5 x=#1 y=#2}\hspace*{#3}%
\protect\myspecial{bmp:#6 x=#1 y=#2}
{#7}
\ifthenelse{\duecol=0}{\myfigend}{\myfigstarend}
}

\newcommand{\quattrobmpperriga}[8]
{
\ifthenelse{\duecol=0}{\myfig{htb!}}{\myfigstar{htb!}}
\vspace*{#2}\relax\noindent
\ifthenelse{\duecol=0}{\hspace*{\duecolhindent}}{}%
\protect\myspecial{bmp:#4 x=#1 y=#2}\hspace*{#3}%
\protect\myspecial{bmp:#5 x=#1 y=#2}\hspace*{#3}%
\protect\myspecial{bmp:#6 x=#1 y=#2}\hspace*{#3}
\protect\myspecial{bmp:#7 x=#1 y=#2}
{#8}
\ifthenelse{\duecol=0}{\myfigend}{\myfigstarend}
}

\newcommand{\bsplit}{\begin{split}}
\newcommand{\esplit}{\end{split}}
\newcommand{\vetz}{\vet{z}}
\newcommand{\vetzh}{\hat{\vet{z}}}

\newcommand{\phor}{p_{H}}
\newcommand{\psihor}{\varphi^{(\pi/2)}}

\newcommand{\pver}{p_{0}}

\newcommand{\psiver}{\varphi^{(0)}}

\newcommand{\pdia}{r_{D}}
\newcommand{\psidia}{\varphi^{(D)}}

\newcommand{\myEy}{\lmathcal{E}_y}
\newcommand{\myEz}{\lmathcal{E}_z}
\newcommand{\myEzthp}{\lmathcal{E}_{z^{(\vartheta_p)}}}
\newcommand{\myN}{N}
\newcommand{\myuid}{s}

\newcommand{\myutre}[3]{\myuid\left(#1,#2;#3\right)}
\newcommand{\myutrerot}[3]{\myuid^{(\vartheta_p)}\left(#1,#2;#3\right)}
\newcommand{\myquad}[3]{#1\left[#2,#3\right]}

\newcommand{\myd}{d}

\newcommand{\mydap}[1]{\myd_{#1}}
\newcommand{\mydeltat}{\delta t}
\newcommand{\myL}{L}
\newcommand{\myB}{B}
\newcommand{\myTp}{T_p}
\newcommand{\myTc}{T_c}

\newcommand{\myTf}{T_f}
\newcommand{\myp}{p}

\newcommand{\myPhi}{\vet{\Phi}}

\newcommand{\myI}{N_1}
\newcommand{\myJ}{N_2}

\newcommand{\RCSM}{RCSM\@\xspace}

\newcommand{\gseteps}[4]{\parbox[c]{#2}{\hspace*{#1}\relax\includegraphics[width=#2,height=#3]{#4}}}
\newcommand{\gsetbmp}[4]{\parbox[c]{#2}{\hspace*{#1}\vskip#3\relax\special{bmp:#4 x=#2, y=#3}}}

\newcommand{\mycollagetwo}[5]
{
\protect\myfig{htb!}
\centering
\begin{tabular}{cc}
 \gsetbmp{0cm}{#1}{#2}{#3}
&\gsetbmp{0cm}{#1}{#2}{#4}
\protect\caption{{#5}}
\vspace*{-0.5em}
\myfigend
}

\def\centereps#1#2#3{\vskip#2\relax\centerline{\hbox to#1{\special
  {eps:#3 x=#1, y=#2}\hfil}}}

\def\centerbmp#1#2#3{\vskip#2\relax\centerline{\hbox to#1{\special
  {bmp:#3 x=#1, y=#2}\hfil}}}

\usepackage{graphicx}

\renewcommand{\centereps}[3]{
\vspace*{1em}\par\noindent
\relax\centerline{\hbox to#1{
	\includegraphics[width=#1,height=#2]{#3}%
	}}}

\date{}
\newcommand{\titolomark}{ }

\title{
Efficient Compressive Sampling  of Spatially Sparse Fields in Wireless Sensor Networks } 

\author{\ifthenelse{\duecol=1}{Stefania Colonnese, Roberto Cusani,  Stefano Rinauro, Giorgia Ruggiero, Gaetano Scarano 
\ifthenelse{\duecol=0}{\thanks{Manuscript submitted ..., revised... .}
 \thanks{S. Colonnese, R.Cusani, S.Rinauro, G. Scarano are with Dip. DIET, Università ``La Sapienza'' di Roma,  
via Eudossiana 18, 00184 Roma, Italy.}
\thanks{e-mail: \{colonnese,rinauro,scarano\}@infocom.uniroma1.it.}
}{}
}
{
Stefania Colonnese,  Roberto Cusani, Stefano Rinauro,  Giorgia Ruggiero, Gaetano Scarano*{\S\footnote{\S~Corresponding author.}} 
\protect\vspace*{1ex}\protect\\ 
\parbox[t]{65ex}{ \protect\centering      \protect\small
                *Dip. DIET,
                 Universit\'a  ``La Sapienza'' di  Roma,
                  via Eudossiana 18, I-00184 Roma, Italy\protect\\
                  Tel:+39.06.44585.500, Fax:+39.06.4873.300\protect\\
                  {\small\texttt{e-mail:~(colonnese,rinauro,scarano)@infocom.uniroma1.it}}
   }
  } 
}

\begin{document}
\maketitle
\begin{abstract}
Wireless sensor networks (WSN), \ie networks of autonomous, wireless sensing nodes spatially deployed over a geographical area, are often faced with acquisition of spatially sparse fields. 
In this paper, we present a novel bandwidth/energy efficient CS scheme for acquisition of spatially sparse fields in a WSN. The paper contribution is twofold. Firstly, we  introduce a sparse, structured CS  matrix and we analytically show that it allows accurate reconstruction of bidimensional spatially sparse signals, such as those occurring in several surveillance application. Secondly, we analytically evaluate the  energy and bandwidth consumption of our  CS scheme   when it is applied to data acquisition in a WSN.  Numerical results demonstrate that our  CS scheme achieves  significant energy and bandwidth savings wrt state-of-the-art approaches when employed for sensing a spatially sparse field by means of a  WSN.\end{abstract}

\begin{keywords}
Compressive sampling, sensor network. 
\end{keywords}
\ifthenelse{\duecol=0}{\bigskip\par\noindent{\small SP-EDICS: SPC-SYNC (Acquisition, synchronization and tracking) }}{}%

\ifthenelse{\duecol=0}{\newpage}{}%
\section{Introduction}
\label{sect:Introduction}%

Wireless sensor networks (WSN) consist of autonomous, cooperative sensors spatially deployed over a geographical area, with applications ranging from surveillance \cite{colglob2012} 
and localization systems \cite{Ghata11}, \cite{Sheu10}, to environmental monitoring for physical field sensing and disaster prevention \cite{Aziz11}, \cite{carli12}. 
WSN  nodes typically acquire the data and communicate them to a  node named Fusion Center (FC), which stores the sensors' readings or forwards them through wired network infrastructures for further processing.

The availability of energy efficient algorithms for data gathering  towards the FC is particularly relevant when the monitoring network is deployed on a large geographical area (e.g. a forest), where highly  efficient routing protocols are required for  sustainable network life-time. Energy efficiency 
is also relevant in those environments where battery recharge or substitution may be unworkable (e.g. in underwater networks) \cite{Ren11},\cite{Hamouda11}. On the other hand, efficient exploitation of the available bandwidth is  an important concern in bandwidth-limited  or interference-limited environments.

The unifying sampling and compression approach of Compressive Sampling (CS) \cite{Bar10} is definitely well-suited to  resource limited WSNs' applications. CS based techniques for energy efficient WSN data gathering have been recently investigated,  with particular reference to the trade-off between the  reconstruction accuracy and the data gathering cost \cite{Quer09}. A highly  efficient approach is provided by Random Sensing (RS),  where at each observation time only a randomly drawn subset of sensors acquires data and  transmits them   to the FC, typically using single hop links. From a theoretical point of view, RS and CS share conditions and procedures for signal reconstruction.
An energy and bandwidth efficient RS procedure appears in \cite{Fazel11}. 

WSN monitoring applications are often faced with acquisition of spatially sparse signals. A typical example is that of temperature monitoring sensor networks for anomalous event (\eg fire)  detection: in the early stages of abnormal system behavior, in which the event is hopefully  detected, the  field is characterized by one or more small spots at levels largely different from the surroundings, and  it can be modeled as a spatially sparse signal. 
RS schemes, such as those analyzed in \cite{Fazel11}, poorly perform in sampling signals that are naturally sparse in the spatial domain, since the actual number of measurements required to reconstruct the field increases and the RS bandwidth/energy efficiency deteriorates. 

This paper  successfully addresses the efficient compressive sampling  of spatially sparse signals in a WSN. We introduce a novel CS scheme that can be realized in a WSN by distributed and parallel data gathering schemes,  with restrained energy and bandwidth consumption for inter-sensor signaling. 
Specifically,  the main contributions of this work are the followings:
\begin{itemize}
\item we introduce a novel CS matrix, and we analytically demonstrate that it satisfies the CS conditions for sparse signal reconstruction;
\item we analytically evaluate the performance, in terms of energy and bandwidth efficiency, of a WSN data gathering scheme based on the herein presented CS matrix; 
\item we show that, on spatially sparse fields, our CS scheme outperforms selected RS and CS state-of-the-art ones in terms of both energy and bandwidth efficiency.
\end{itemize}

The remainder of the paper is organized as follows: Sect.\ref{sect:model} describes  the WSN system model, Sect.\ref{sec:cs} briefs the CS basics, and Sect.\ref{sect:relwork} discusses related works on CS/RS acquisition in WSN. In Sect.\ref{sect:radonCS} we illustrate our original CS scheme and  in Sect.\ref{sec:RLWSN} we  apply it in a WSN and we analyze the related energy consumption and bandwidth occupancy.  Sect.\ref{sec:numres} presents numerical results assessing that our CS scheme outperforms of state-of-the-art approaches in terms of bandwidth/energy efficiency. Sect.\ref{sec:concl} concludes the paper.

\section{System Model and Network Scenario}
\label{sect:model}
Let us consider a physical field 
represented by a  bi-dimensional time varying signal \(\myutre{x}{y}{t}\), \((x,y)\in\mathcal{R}^2\) monitored through a  a grid of \(N=\myI\times\myJ\) sensors deployed over a bi-dimensional covered area, being \(\myI\) and \(\myJ\)  the number of sensors  distributed in the horizontal and vertical dimension, respectively.
A selected sensor collects the data from the others and plays the role of  FC. 
An example of such a WSN scenario is depicted in Fig.\ref{fig:net}: the FC is placed at the center of the network, and each sensor is placed at distance \(\mydap{k}, k=1,\dotsc,N-1\) from the FC.

\myfig{t!}
\centereps{5cm}{3.78cm}{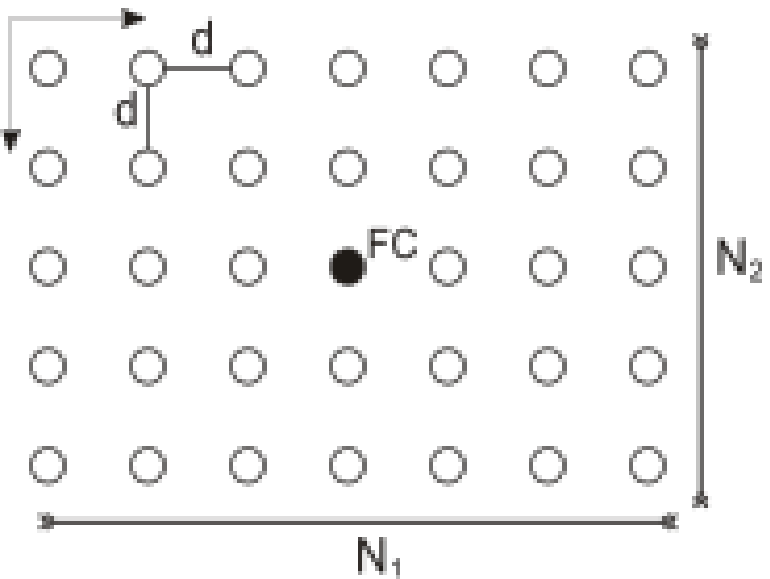}
\caption{Sensor network scenario.}
\label{fig:net}
\myfigend


The sensors  periodically  measure  \(\myutre{x}{y}{t}\) and transmit their readings to the FC for monitoring. 
The sensing period \(\mydeltat\) is selected to be almost equal to the coherence time \(\myTc\) of \(\myutre{x}{y}{t}\)%
\footnote{  The coherence time \(\myTc\) is defined as the time interval over which the process almost de-correlates in time.}.
At time \(t_k=t_0+\mydeltat\), the sensor at the location \((n_1\myd, n_2\myd)\)   acquires the noisy measurement:
\begin{equation}
\myquad{z}{n_1}{n_2}=\myutre{n_1\myd}{n_2\myd}{t_k} 
\label{eq:noisysignal}
\end{equation}
where, for simplicity sake, we have dropped the temporal variable and we have set the horizontal and vertical inter-sensor distances equal to \(\myd\). 
Then, the sensors implement a suitable dissemination protocol to forward the measured value to the FC.
The FC collects all the data from the sensors so as to reconstruct a representation  of the overall field \(\myquad{z}{n_1}{n_2}\).

Efficient  data dissemination towards the FC is widely debated in the literature, since energy efficiency  affects the  network lifetime, 
especially relevant in scenarios where the deployment of the sensors is difficult or expensive, 
whereas bandwidth efficiency enables WSN monitoring in bandwidth-limited media, e.g. underwater, 
or in geographical areas where different sensor networks coexist.
CS and RS paradigms provide a theoretical framework for highly efficient field monitoring, provided that the monitored data are sparse in a suitable domain. 
We briefly recall in Secs.\ref{sec:cs}, \ref{sect:relwork} the basics of CS  and the related work on CS and RS  application in WSN, respectively.


\section{An Introduction to Compressive  Sampling}
\label{sec:cs}
Let us compactly represent the  bidimensional sequence \(\myquad{z}{n_1}{n_2}\) 
by the lexicographically ordered vector \(\vetz\):
\[
\vetz=\left[\myquad{z}{1}{1} \cdots \myquad{z}{1}{\myJ}\;\myquad{z}{2}{1}\cdots  \myquad{z}{\myI}{\myJ}\right]\ttt
\]
The vector \(\vetz \in \mathbb{R}^N\)  is sparse if  the number of its non-zero samples is restrained compared to its own dimension \(N\); rigorously, \(\vetz\)  is said to be \(K\)-sparse if the number of non-zero components is \(K\), either in the spatial or in a transformed domain (\eg Fourier, wavelet, etc.). 
Compressive sampling  (CS) provides a 
framework for  sensing and compression of a sparse signal.

According to the CS paradigm, compression of sparse signals is performed jointly with the acquisition.
Specifically, \(\vetz\) is represented by \(M\) projections defined as follows:
\begin{equation}
\vety=\myPhi\vetz
\label{eq:ycs}
\end{equation}
being \(\myPhi\) a suitable \textit{measurement matrix}  of size \(M\times N\).

A fundamental outcome of CS is  that, under suitable conditions on the sensing matrix \(\myPhi\),  if \(\vetz\) is \(K\)-sparse  it can be accurately recovered from  the projections in \(\vety\) provided that \(K < M < N\). 
Specifically, 
the sensing matrix \(\myPhi\) 
must satisfy the so called \textit{Restricted Isometry Property} (RIP), \ie, given a constant \(\delta\), for all \(K\)-sparse signals \(\vetz\) it must stand: 
\[
(1-\delta)||\vetz||_2^2\le||\myPhi\vetz||_2^2\le(1+\delta)||\vetz||_2^2
\]

It is proved \cite{CandesSPM} that, for values of \(\delta\) small enough, sparse signals can be perfectly recovered from compressive sensing measurements. 
%
In \cite{bar09},  the authors show that  the RIP property is verified when the measurements energy \(||\Phi_R \vetz||^2\) is more and more concentrated, in probability, around the value \(||\vetz||_2^2\) as far as the number of measurements increases.

Reconstruction can be  achieved either by  solving the following optimization problem
\begin{equation}
\hat{\vetz}=\arg\min_{\vet{t}}||\vet{t}||_1\hspace*{3ex}\text{s.t.}\hspace*{3ex}\vety=\myPhi\vet{z}
\label{eq:csl1}
\end{equation}
or by a greedy iterative pursuit of the support of \(\vetz\); examples of this latter approach are provided by the orthogonal matching pursuit (OMP) \cite{Tropp07} and the compressive sampling matching (CoSaMP) algorithms \cite{Needell08}. 

In \eqref{eq:ycs} we recognize that randomly sampling the components of \(\vetz\) and collecting them in \(\vety\) is equivalent to realize CS using a particular sensing matrix; therefore, many CS theory results apply to RS, too. Both RS and CS techniques have been considered for application in WSNs.

\section{Related Works}
\label{sect:relwork}

Efficient  sensing and data gathering in a WSNs by means of CS and RS based techniques has  aroused lively interest in recent literature. 
In  \cite{Bajwa06}, a CS based distributed, communication architecture   is exploited to minimize the latency for information retrieval under a favorable power-distortion trade-off whereas in \cite{Shen08} a  CS based sensing and data gathering procedure is analyzed for the case of network routing tree topologies. 
In \cite{Caione12}, maximization  of large scale WSN  lifetime is pursued  by means of a fully distributed algorithm according to which each sensor autonomously performs classical or compressive sampling in order to reduce the number of transmitted packets. 

In \cite{Quer09} the authors analyze a  RS and  multi-hop data gathering scheme.
In this  scheme only randomly selected  nodes measure the field and transmit their readings to the FC through the 
specific multi-hop paths. 
While a sensor reading is  
routed towards the FC, its value is combined with the ones sensed by the sensors in the path, so that 
each random projection provided to the FC is  built by accumulating randomly weighted sensors readings along a network path. Energy efficiency is pursued if  the number of nodes contributing to each projection is low.

In \cite{Luo10} a peculiar form of the CS sensing matrix is proved to exhibit good reconstruction properties while still being able to reduce the number of inter-sensors transmissions. The structure of the sensing matrix, originally designed for WSNs with chain topology is viable of an extension to more complex scenarios, provided suitably tree-structured routing paths are designed from the exterior of the network towards the FC. 

Recently, in \cite{Fazel11} a RS based  sensor network framework for underwater systems has been introduced.
Differently from the works in \cite{Caione12}, \cite{Quer09}, \cite{Shen08}, a single hop network is considered, where the sensors directly communicate with the FC. 
Every  sampling interval \(\mydeltat\), each sensor senses the field and transmits it directly to the FC with probability \(\myp\). 
The sensing probability \(\myp\) is suitably chosen  in order to let the FC acquire sufficiently many data for field reconstruction.
The approach in \cite{Fazel11}   favorably merges the RS procedure with a random access protocol, thus obtaining a significant reduction in both the consumed energy and the occupied bandwidth.

The main reasons why the above discussed methods achieve significant performances in efficient use of network resources is either  in the fact that, during a sampling interval, only a sub-set of the sensors measures the field by means of RS, or in the fact that CS acquisition is actually realized jointly with the routing procedure. 

Recent studies \cite{Shen08}, \cite{Luo10}, suggest that in case of spatially sparse signals   the energy/bandwidth of RS and CS efficiency deteriorates, since
reconstruction accuracy is guaranteed only when a large number of sensors contribute to the measurements. 
In \cite{Shen08}, the authors remark how it is difficult to design a RS matrix suited to  sparse signals
and still allowing  an efficient network routing.  In \cite{Luo10} it is pointed out that, on spatially sparse signals, CS techniques still guarantee reconstruction accuracy but at an increased number of measurements (\eg \(M\) up to \(50\%\) of \(N\)) whereas, for the same values of \(M\) and \(N\), RS only opportunistically achieves reconstruction.

In several WSN applications,  the sensed field   indeed contains local fluctuations and abnormal readings, and it is well modeled as a spatially sparse signal.  This motivated us to concern ourselves with the design of a sensing matrix suited to spatially sparse signals, as described in the following Section.

\renewcommand{\pver}{y^{(0)}}
\renewcommand{\pdia}{y^{(\pi/4)}}
\renewcommand{\phor}{y^{(\pi/2)}}
\renewcommand{\psidia}{\varphi^{(\pi/4)}}

\unepsperriga{7.89cm}{6cm}{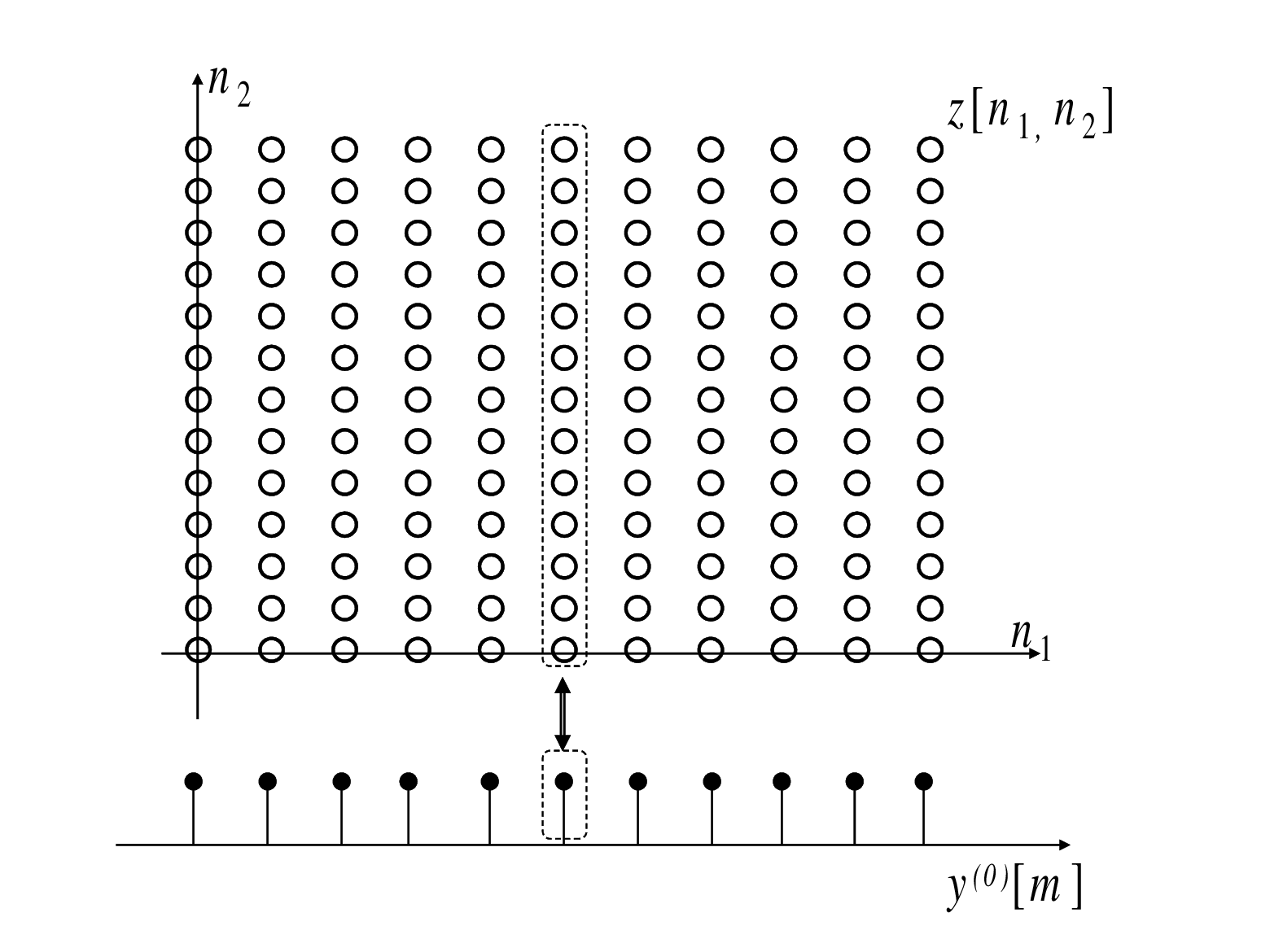}{\vspace{-1em}\caption{Bidimensional sequence \protect\(\myquad{z}{n_1}{n_2}\): column-wise accumulation for computation of the measurements \protect\(\pver[m]\protect\) \label{fig:ssg}}}

\section{CS using Radon-like random projections}
\label{sect:radonCS}

In a nutshell, we aim at devising a sensing matrix that 
represents a spatially sparse field \(\vetz\) in a domain  such that dropping \(N\!-\!M\) components does not prevent signal reconstruction.

Recalling that the Radon Transform has the dual properties of i) compressing spatial domain straight lines into transform domain pulses \cite{myjasp}, ii) expanding spatial discrete pulses into as many nonzero values as the number of considered Radon projections \cite{flusser},   we recognize that the Radon Transform provides a mean for redundant representation of sparse signals built by spatially isolated pulses. 
Thereby,
we resort to a spatially sparse signal CS scheme inspired to the Radon projections computation.
\newcommand{\mydiagafirst}{\begin{array}{cccc}
\psiver_{0}[0]&0      &\cdots&0\\
0   &\psiver_{1}[0]   &\cdots&0\\
0   &\vdots &\ddots&0\\
0   &0      &\cdots&\psiver_{N_2-1}[0]
\end{array}}
\newcommand{\mydiagalast}{\begin{array}{cccc}
\psiver_{0}[N_1-1]&0      &\cdots&0\\
0   &\psiver_{1}[N_1-1]   &\cdots&0\\
0   &\vdots &\ddots&0\\
0   &0      &\cdots&\psiver_{N_2-1}[N_1-1]\phantom{-1}
\end{array}}
\newcommand{\mydiagbfirst}{\begin{array}{cccc}
\phantom{-1}\psihor_{0}[0]&\psihor_{0}[1]&\cdots&\psihor_{0}[N_2-1]\\
0   &0   &\cdots&0\\
0   &0   &\dots &0\\
0   &0   &\cdots&0
\end{array}}
\newcommand{\mydiagblast}{\begin{array}{cccc}
0   &0   &\cdots &0\\
0   &0   &\cdots &0\\
0   &0   &\cdots &0\\
\psihor_{N_1-1}[0]&\psihor_{N_1-1}[1]&\cdots&\psihor_{N_1-1}[N_2-1]
\end{array}}
\newcommand{\mydiagdfirst}{\begin{array}{cccc}
0    &\cdots &0       &\phi\\
0    &0      &\phi    &0\\
0    &\phi   &\cdots  &0\\
\phi &0      &\cdots  &0\\
0    &0      &\cdots  &0\\
0    &0      &\cdots  &0\\
0    &0      &\cdots  &0
\end{array}}
\newcommand{\mydiagdlast}{\begin{array}{cccc}
0    &0    &\cdots  &0\\
0    &0    &\cdots  &0\\
0    &0    &\cdots  &0\\
0    &0    &\cdots  &\phi\\
0    &0    &\phi  &0\\
0    &\phi    &\cdots  &0\\
\phi &0 &\cdots  &0
\end{array}}

\begin{figure*}
\begin{equation}
\label{eq:psiver}
\Phi_R^{(0)}=\left[
\mydiagafirst \cdots \cdots \mydiagalast
\right]
\end{equation}
\begin{equation}
\label{eq:psihor}
\Phi_R^{(\pi/2)}=\left[
\mydiagbfirst \cdots \cdots \mydiagblast\right]
\end{equation}
\end{figure*}


To elaborate, let us consider the  bidimensional sequence \(\myquad{z}{n_1}{n_2} \) of finite size \(\myI\times \myJ\)and let us present few examples of  measurements computed  in analogy to the Radon projections.

First, let us consider the column-wise accumulation of a randomly weighted version of  \(\myquad{z}{n_1}{n_2}\): 
\begin{equation}
\pver[m]=
\sum_{i=0}^{\myJ-1}
\psiver_{m}[i] \myquad{z}{i}{m}, \;m=0, \dotsc, \myI-1
\label{eq:pver}
\end{equation}
where  \(\psiver_{m}[i], i=0, \dotsc N_1-1, m=0, \dotsc, N_2-1 \) are i.i.d. zero mean Gaussian random variables of equal variance \(\sigma^2_{\varphi}\), \ie \(\psiver_{m}[i]
\sim 
\lmathcal{N}(0, \sigma^2_{\varphi})\);
an example of the formation of the measurements \(\pver[m]\) is illustrated in Fig.\ref{fig:ssg}.
The measurements \(\pver[m]\) differ from a Radon projection in that each sample is randomly weighted. Besides, they differ from classical CS measurements in that each measure is obtained only from a subset (namely a column) of the values in \(\myquad{z}{n_1}{n_2}\) rather than on all the samples \(\myquad{z}{n_1}{n_2}\).

Definition of the row-wise random projections of \(\myquad{z}{n_1}{n_2}\) is straightforward, namely:
\begin{equation}
\phor[m]=
\sum_{i=0}^{\myI-1}
\psihor_{m}[i] \cdot\myquad{z}{m}{i}, \;m=0, \dotsc, \myJ-1
\label{eq:phor}
\end{equation}
By analogy, we can define the diagonal-wise projection of \(\myquad{z}{n_1}{n_2}\) as well:
\begin{equation}
\pdia[m]=\!\left\{
\begin{split}
&\sum_{i=0}^{\myI-1-m}
\!\psidia_{m}[i] \myquad{z}{i+m}{i}, \\&\phantom{\psidia_{m}[i]}\;m\!=0,\;\dotsc,\;\!\myI-1
\\&
\sum_{i=0}^{\myJ-1-|m|}
\!\psidia_{m}[i] \myquad{z}{i}{i+m},  \\&\phantom{\psidia_{m}[i]}\;m\!=-\myJ+1, \dotsc, \!-1
\end{split}
\right.
\label{eq:pdia}
\end{equation}

The  expressions \eqref{eq:pver}-\eqref{eq:pdia} represent randomly weighted accumulations of \(\myquad{z}{n_1}{n_2}\) over ridge paths. Let us now generalize the above expressions by regarding them as obtained by column-wise accumulation of a suitably rotated version of the sequence \(\myquad{z}{n_1}{n_2}\).

Let \(\myquad{z^{(\vartheta_p)}}{n_1}{n_2}\) be  a \(\vartheta_p\)-radiant clockwise rotated%
\footnote{Formally, we obtain the \(\vartheta_p\)-radiant clockwise rotated version of the image \(\myquad{z}{n_1}{n_2}\) by regular sampling of the rotated field \[\myutrerot{x}{y}{t}=\myutrerot{x \cos \vartheta_p+y \sin \vartheta_p}{x \sin \vartheta_p-y \cos \vartheta_p}{t}. \] }
version  of the image \(\myquad{z}{n_1}{n_2}\). 
The 
size of \(\myquad{z^{(\vartheta_p)}}{n_1}{n_2}\) varies with \(\vartheta_p\), and we denote as  \(K(p), p=0,\dotsc P-1\) the number of columns of \(\myquad{z^{(\vartheta_p)}}{n_1}{n_2}\).

We  generalize the definitions  \eqref{eq:pver}-\eqref{eq:pdia}  by considering the collection of the random projections of \(\myquad{z^{(\vartheta_p)}}{n_1}{n_2}\) over a finite set of \(P\) directions \(\vartheta_p\):
\begin{equation}
y^{(\vartheta_p)}[m]=\sum_{i}\varphi^{(\vartheta_p)}_{m}[i] \cdot\myquad{z^{(\vartheta_p)}}{i}{m} \label{eq:rlike}
\end{equation} 
for   \(m=0,\dotsc K(p)-1\),  \(p=0,\dotsc P-1\). 
Since the accumulation in  \eqref{eq:rlike} recalls
 the column-wise accumulations employed in the computation of the discrete Radon transform, we refer to the projections  in \eqref{eq:rlike}   as \textit{Radon-like} random projections. 

Let us now collect the  \(P\) Radon-like random projections:
 \[\vety^{(\vartheta_p)}=\left[y^{(\vartheta_p)}[0]\cdots y^{(\vartheta_p)}[K(p)-1]\right]\!\tras\!, p=0,\dotsc P-1\]
in a measurement vector \(\vety\):
\[\vety=[ \left.\vety^{(\vartheta_0)} \right.\tras\cdots \left.\vety^{(\vartheta_{P-1})} \right.\tras]\tras\] 

The measurements \(\vety\) are computed from \(\vetz\) as:
\begin{equation}\vety=\Phi_R \vetz\label{eq:radon}\end{equation} using the \(\myI\times \myJ\) random sampling matrix \(\Phi_R\) defined as:
\[\Phi_R=\left[\begin{array}{cccc}
\Phi_R^{(\vartheta_0)}\\ \hdashline
\vdots\\ \hdashline
\Phi_R^{(\vartheta_{P-1})}\\ 
\end{array}\right]\]
where \(\Phi_R^{(\vartheta_{p})}, p=0,\dotsc P-1\) are the suitably defined sparse random matrix realizing the accumulation in \eqref{eq:phor}-\eqref{eq:pdia}.
For the sake of clarity, we report the matrices  \(\Phi_R^{(0)}\) and \(\Phi_R^{(\pi/2)}\), corresponding to the horizontal and the vertical projection, in \eqref{eq:psihor} and \eqref{eq:psiver} respectively.

Let us remark that, even though the samples contributing to each \(\vety\) component are the same that would have contribute to a specific  Radon projection of  \(\vetz\),  due to the random weighting   the measurements vector \(\vety\) is not -and it is not even easily related to-  the Radon transform of \(\vetz\).  

In the followings, we demonstrate that the conditions given by the CS theory for reconstructing  \(\vetz\) from  \(\vety\) stand.
Before turning to mathematics, let us observe that if the image \(\myquad{z}{n_1}{n_2}\)  is built by sparse isolated pulses, each pattern contributes, a part a suitable randomly weighting, to each one of the \(P\) Radon-like random projections \(y^{(\vartheta_p)}[m],  \;m=0,\dotsc K(p)-1, \;p=0,\dotsc P-1\). Thereby, the set of \(P\) Radon-like random projections is a \(P\)-redundant representation of the pulse. 
This motivated us to formally demonstrate that the Radon-like random projections \(y^{(\vartheta_p)}[m],  \;m=0,\dotsc K(p)-1, \;p=0,\dotsc P-1\) satisfy the conditions of a CS measurement set.


\subsection{Restricted Isometry Property of the Radon-like sampling matrix}
To formally state that the above introduced simplified sampling structure is feasible for accurate field reconstruction, we  shall demonstrate that the random measurements \(\vety\) evaluated as n \eqref{eq:radon} are consistent CS measurements, \ie they substantially preserve the information of the sampled sequence \(\myquad{z}{n_1}{n_2}\).

Formally,  we need to prove that  the sampling matrix \(\Phi_R\) satisfies the condition known as Restricted Isometry Property (RIP). Specifically, 
let us denote by \(\myEz\ggdef|| \vetz||^2\)  the quadratic norm of the vector \(\vetz\). If the matrix \(\Phi_R\) satisfies:
\begin{equation}
\label{eq:rip}
(1-\delta)\;\myEz\leq||\Phi_R \vetz||^2\leq(1+\delta)\;\myEz
\end{equation}
with high probability, %
then, any sparse sequence \(\vetz\) can be perfectly recovered from CS measurements \(\vety=\Phi_R \vetz\) \cite{CandesSPM}.

The RIP in \eqref{eq:rip} asserts
 that the measurements energy \(\myEy\ggdef||\Phi_R \vetz||^2\) is strongly concentrated  around the value \(\myEz\). 
Preliminary results on the RIP property of a Radon-like CS matrix appear in \cite{colisccsp12}. In \ref{app:proof}, we extend these results and, following the approach in \cite{bar09},
we  show that the following property stands. 

\textbf{Property 1:} If   the nonzero  entries in \(\Phi_R\) are i.i.d. zero mean Gaussian random variables with equal variance \(\sigma^2_{\varphi}=1/P\), the following  concentration inequality stands:
\begin{equation}
\label{eq:ci}
\Pr\{|\myEy-\myEz|\geq\delta\}\leq \epsilon
\end{equation}
provided that \(P\geq 2 K^2 C^2_2 \log(2/\epsilon)/\delta^2 \) being \(C_2\) a suitable constant.
From Property 1, the RIP property in \eqref{eq:rip} immediately follows.

\subsection{Further Remarks}

The sampling matrix \(\Phi_R\) differs from the full sampling matrices  usually adopted in CS since it is  sparse. In the following, we show how the sparsity of the Radon-like sensing matrix \(\Phi_R\) can be leveraged on to significantly simplify the CS measurements computation  in a WSN,  so as to  reduce the consumed energy and employed bandwidth.

The definition of  \(y^{(\vartheta_p)}[m]\) in  \eqref{eq:rlike} given above is consistent and useful from an analytical point of view. When turning to evaluation of \(y^{(\vartheta_p)}[m]\) in a WSN, the evaluation of \(\myquad{z^{(\vartheta_p)}}{n_1}{n_2}\) is not accomplished, and the measurements computation is realized throughout the data gathering stage. 

Finally, the concentration inequality  \eqref{eq:ci} guarantees that the sequence \(\vetz\) can be recovered from the measurements \(\vety\) with high probability, as far as the number \(P\) of considered projections increases.  
Since
CS convergence is assured only in probability, 
the CS measurement experiment could be repeated.  
In a WSN,  data are periodically sensed and routed to the FC, and a small probability of mis-reconstruction can be tolerated since it can be recovered in the subsequent sampling interval. Furthermore, integration of independently drawn measurements acquired in a WSN during different temporal intervals can be performed at the FC. This interesting research issue is deferred to further studies.

\newcommand{\vlength}{8.0cm}
\newcommand{\hlength}{8.0cm}
\newcommand{\vlengtha}{8.cm}
\newcommand{\hlengtha}{8.cm}

\section{Radon-like CS in a WSN}
\label{sec:RLWSN}
\newcommand{\myP}{P}
\newcommand{\myNs}{N_{\text{slot}}}

In a WSN application scenario, the  \(\myP\) Radon-like projections correspond to randomly weighted sums of the values sensed by different sub-sets of WSN sensors. The sums can be evaluated using  different techniques. 

Following the approach in \cite{Haput08}, the sensors within a subset can synchronously transmit their weighted sensed values, and the sum can be realized at the FC by on air analogical superimposition of received signals. This protocol requires  a strict control of the power received by the FC from each sensor. Precisely, each sensor node needs to estimate the channel seen towards the FC, in order to pre-compensate the transmitted value according to the channel attenuation. Thereby, although feasible in principle, this approach requires a sophisticated processing and a tight power control by the sensor nodes.

According to a data gathering paradigm, the projections are computed within the network by sub-set of sensors while they are forwarding their sensed values to the FC. 
The sums in  \eqref{eq:rlike} can then be realized by routing and accumulating values of \(\myquad{z}{n_1}{n_2}\) over suitably tilted paths in the network grid discrete support. 
Here we refer to such a data gathering approach, and we infer some consequences  from the peculiar sparse structure of the matrix \(\Phi_R\) on the  computation procedure.

Firstly, we observe that in every row, the non-zero coefficients of the matrix \(\Phi_R\) are arranged so as to obtain \(y^{(\vartheta_p)}[m]\) as the sum of the values of a column of the rotated image 
\(\vet{z}^{(\vartheta_{p})}\). 
When collecting the measurement in a WSN,  each projection \(y^{(\vartheta_p)}[m]\)   can then be computed by accumulating measurements throughout a specific, suitably tilted, grid path.
Secondly, we observe that in every column of the matrix \(\Phi_R\)  there are only \(P\) non-zero values. Hence, each value \(\myquad{z}{i}{j}\) shall contribute only to \(P\) out of \(M\) projections \(y^{(\vartheta_p)}[m]\). When realizing the Radon-like CS in the  sensor network grid, each sensor shall transmit its value \(P\) times.

Based on these premises, we recognize that the sparse structure of the matrix \(\Phi_R\) results into two main features of   Radon-like projections computation in a WSN:
\begin{itemize}
\item  the computation of  each projection \(y^{(\vartheta_p)}[m]\) is   performed in a distributed way within the WSN, and it requires  signaling among grid sensors which are adjacent along a WSN path; 
\item
 the accumulation along different paths can then be realized in parallel, provided that the distance between contemporaneously signaling nodes is kept large enough. 
\end{itemize}

We now turn to quantifying the  energy consumption and bandwidth occupancy required for realizing the Radon-like CS by means of a data gathering procedure in a WSN. 
In \ref{subsec:RLperf} we evaluate  the energy consumption and bandwidth occupancy of the Radon-like CS scheme in a WSN. 
Next, we compare these results with selected state of the art schemes, namely  those referring to the conventional network and to the Random Sensing  approaches described in Sect.\ref{sect:Introduction}, respectively detailed in subsections \ref{sub:conv} and \ref{sub:racs}.

\newcommand{\myItilde}{\tilde{\myI}}
\newcommand{\myJtilde}{\tilde{\myJ}}
\newcommand{\mytaui}{\nu_0}
\newcommand{\myto}{t_0}
\newcommand{\mytf}[1]{t^{(#1)}_f}
\newcommand{\myK}{K}

\newcommand{\myPk}{P^{(k)}}
\newcommand{\myEk}{E^{(k)}}
\newcommand{\myG}{G}
\newcommand{\myEconv}{E_{\text{CON}}}
\newcommand{\myBconv}{B_{\text{CON}}}
\newcommand{\myErr}{E_{\text{RR}}}
\newcommand{\myBrr}{B_{\text{RR}}}
\newcommand{\myErd}{E_{\text{RD}}}
\newcommand{\myBrd}{B_{\text{RD}}}
\newcommand{\myErad}{E_{\text{RL}}}
\newcommand{\myBrad}{B_{\text{RL}}}
\newcommand{\myfloor}[1]{\lfloor#1\rfloor}
\newcommand{\myceil}[1]{\lceil#1\rceil}
\newcommand{\myal}{\alpha}
\newcommand{\myTpCon}{\myTp^{(\text{CON})}}
\newcommand{\myTprr}{\myTp^{(\text{RR})}}
\newcommand{\myTprd}{\myTp^{(\text{RD})}}
\newcommand{\myTprad}{\myTp^{(\text{RL})}}
\newcommand{\myps}{p_s}
\newcommand{\myqs}{q_s}
\newcommand{\myEsh}{E^{*}}
\newcommand{\mygammap}{\gamma_P}
\newcommand{\mydeltap}{\delta_P}

\subsection{Radon-Like CS efficiency}
\label{subsec:RLperf}

We  now evaluate the allocated bandwidth and consumed energy for entirely collecting the measurements  in a time \(\myTc\) in the WSN scenario in Sec.\ref{sect:model}. The actual bandwidth occupation and energy consumption depend not only on  the WSN structure but also  on the adopted data gathering procedure. 
Without loss of generality,  we refer to the suboptimal data gathering algorithm sketched out in  \ref{app:dga}, and here briefly summarized for the reader's convenience.

According to the algorithm in  \ref{app:dga}, the sensors transmit their readings to the FC by data gathering through suitable multi-hop paths. Figs.\ref{fig:datagath}-\ref{fig:diag} illustrate, as an example, the multi-hop paths selected for the computation of the projections \(\phor[m], \pdia[m]\) within a network quadrant.

\unepsperriga{9cm}{6cm}{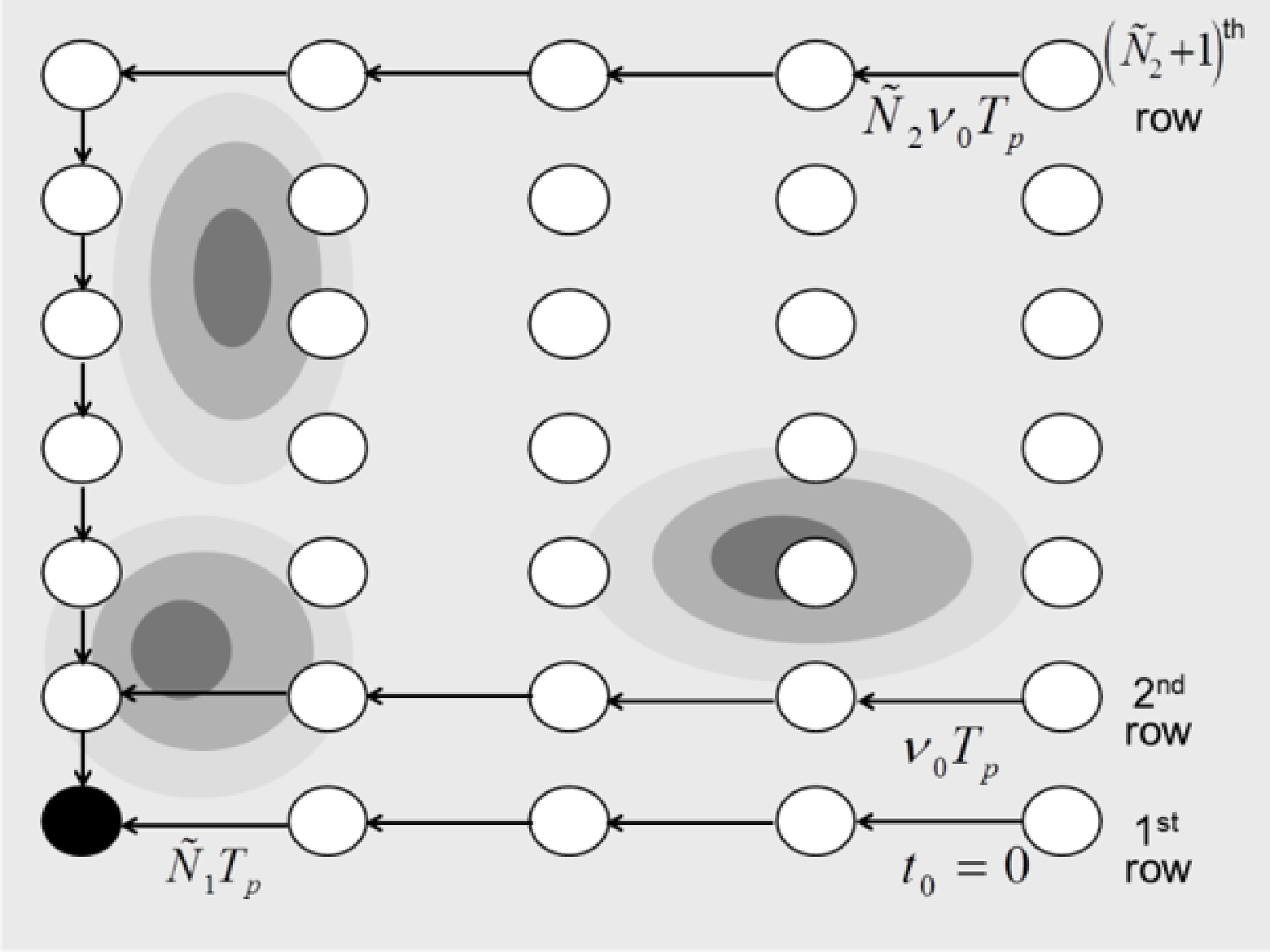}{\caption{Data gathering algorithm: spatial organization of the sensors' transmissions for horizontal projections evaluation in the first quadrant. Time stamps indicate when the node starts transmitting.\label{fig:datagath}}}
\unepsperriga{9cm}{6cm}{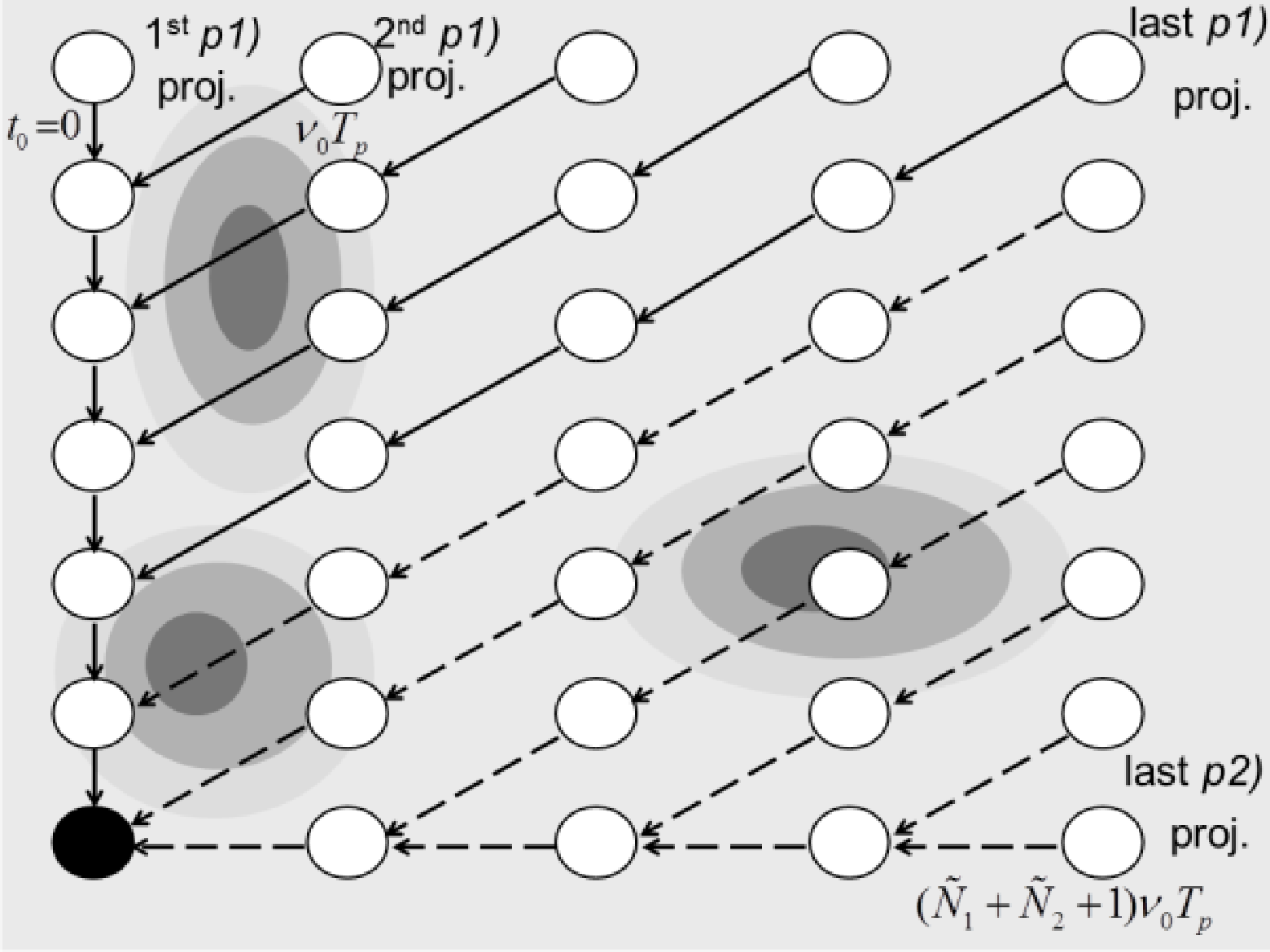}{\caption{Data gathering algorithm: spatial organization of the sensors' transmissions for diagonal projections evaluation in the first quadrant. Solid arrows indicate \textit{p1)} paths, dashed arrows indicate \textit{p2)} paths. Time stamps indicate when the node starts transmitting. \label{fig:diag}}}

A deterministic (collision free) Time Division Multiple Access (TDMA) is adopted. Signaling takes place between adjacent nodes only. Parallel transmission of sufficiently apart nodes are considered.

In  \ref{app:dga}, we evaluate two parameters that directly affect the energy and bandwidth consumption of the data gathering algorithm, namely total number of single hop transmissions  \(N_{\text{TX}}\), and the total number \(N_{\text{TS}}\) of time slots needed to collect all the sensors' readings to the FC.

The total number of node transmission  \(N_{\text{TX}}\) for the algorithm under concern comes out to vary linearly with the network size:
\begin{equation}
N_{\text{TX}}\approx\mygammap\cdot N,
\end{equation}
being \(\mygammap\) a scalar factor depending on the adopted \(P\) projections. The guidelines for calculating \(\mygammap\) are given in the \ref{app:dga} where we also evaluate the value of \(\mygammap\) for different sets of directions \(\vartheta_p\).

Based on the same data gathering algorithm, we have evaluated the number of time slots needed to collect all the sensors' readings to the FC. The number of time slots turns out to vary just with the square root of the network size, that is:
\begin{equation}
N_{\text{TS}}\approx\mydeltap\cdot \sqrt{N},
\end{equation}
being  \(\mydeltap\) a factor depending on the considered projection directions \(\vartheta_p\). \ref{app:dga} reports the guidelines for the evaluation of the parameter \(\mydeltap\) for different sets of directions \(\vartheta_p\).

We observe that, as a consequence of the sparsity of the Radon-like sensing matrix, i) the number of transmission varies only linearly with the network size \(N\); ii)  being the algorithm parallelized, the number of time slots varies linearly with the square root of the network size \(N\).

With these results, we are able to evaluate the allocated bandwidth and consumed energy for entirely collecting the measurements  in a time \(\myTc\) in the WSN scenario described in Sect.\ref{sect:model}.

Projection evaluation according to the data gathering scheme detailed in \ref{app:dga} accounts for a series of transmissions among neighboring nodes. 
The  energy spent for a single-hop  transmission is given by \(
E_{\text{SH}}=G d^2_{\text{SH}} \myTprad
\)
being \(d_{\text{SH}}=\myd\) or \(\sqrt{2}\myd\) the distance for horizontally, vertically and diagonally adjacent nodes,  and \(\myTprad\) the time needed for packet transmission.

Overall, we can express  the total consumed energy%
\footnote{We herein approximate also the inter-sensor distance in the diagonal direction to \(\myd\).}
 for the Radon-Like CS based approach as 
\begin{equation}
\myErad=N_{\text{TX}}\myG \;\myd^2\myTprad
\end{equation}
being  the overall number of transmission.

Besides, in order to assure that the measurements are collected within a maximum time of \(T_c\), the packet time \(\myTprad\) needs to meet
\[
\myTprad=\dfrac{\myTc}{N_{\text{TS}}}
\]
so that we have
\begin{equation}
\myErad=\dfrac{\mygammap}{\delta(P)}\myTc\myG \;\myd^2 \cdot\sqrt{N}
\label{eq:erad2}
\end{equation}

%

The occupied bandwidth, defined as the packet length in bits \(\myL\) over the packet transmission time, for the Radon-Like approach is finally evaluated as
\begin{equation}
\myBrad=\dfrac{\myL}{\myTprad}=\dfrac{\myL}{\myTc}\mydeltap\cdot \sqrt{N}
\label{eq:brad}
\end{equation}

\subsection{Conventional network efficiency}
\label{sub:conv}
As a benchmark for reconstruction quality and energy/bandwidth consumption comparison purposes, we refer to a conventional network in which, at time \(t\), all of the sensors  sample the underlying field 
and  transmit their measurements via a TDMA scheme.
\(\myN\) time slots are organized in a frame structure of duration   \(\myTf=N\myTp\).
Recalling that the FC needs to reconstruct the sensed field every \(\myTc\), it must be \(\myTf\le\myTc\).

In the conventional network, all the sensors access the medium for packet transmission in a TDMA fashion. Thus, the conventional network packet transmission time \(\myTpCon\), is given by \(\myTpCon=\myTc/N\).
The overall energy consumed  by the conventional network is then evaluated as
\begin{equation*}
\myEconv=\sum_k\myPk\myTpCon=\sum_k\myG\;\mydap{k}^2\myTpCon
\end{equation*}
Since the FC is located at the center of a regular grid, as depicted in Fig.\ref{fig:net}, the value of \(\mydap{k}\) can be expressed in terms of the inter-sensor distance \(\myd\):  
\(\mydap{k}=\myd\;\sqrt{i^2+j^2}\)
where \(i\), \(j\) range in \(\left(-\!\myfloor{\myI/2},\myfloor{\myI/2}\right)\),\(\left(-\!\myfloor{\myJ/2},\myfloor{\myJ/2}\right)\), respectively; then, \(\myEconv\) rewrites as
\begin{equation}
\myEconv=\sum_i\sum_j(i^2+j^2)\myG \;\myd^2\myTpCon
\label{eq:sumi}
\end{equation} 
The sums over \(i\) and \(j\) in \eqref{eq:sumi} 
can be approximated as
\footnote{Recall that \(\sum_{l=1}^{n} l^2 = n(n+1)(2n+1)/6\).}%
\[
\sum_i\sum_j (i^2+j^2)\simeq\left(\dfrac{\myJ\myI^2 + \myI\myJ^2}{48}\right)=\dfrac{\myI\myJ}{48} \left({\myI^2 +\myJ^2}\right)
\]
Denoting \(\myI\ggdef\alpha_1 \sqrt{N}\), \(\myJ\ggdef \alpha_2\sqrt{N}\),
the overall energy consumption for the conventional network sums up to
\begin{equation}
\myEconv=\dfrac{\alpha_1 \alpha_2 (\alpha_1^2 +\alpha_2^2 )}{48}\;\myG \;\myd^2 \myN^2\myTpCon
\label{eq:Econv1}
\end{equation}

By setting \(\myTpCon=\myTc/\myN\) in \eqref{eq:Econv1} we come up with:
\begin{equation}
\myEconv=\dfrac{\alpha_1 \alpha_2 (\alpha_1^2 +\alpha_2^2 )}{48}\;\myG \;\myd^2 \myN\myTc
\label{eq:Econv2}
\end{equation}

The occupied bandwidth for the conventional network is:
\begin{equation}
\myBconv=\dfrac{\myL}{\myTpCon}=\myN\dfrac{\myL}{\myTc}
\label{eq:Bconv}
\end{equation}

\subsection{RS Efficiency}
\label{sub:racs}
We now summarize the energy and bandwidth consumption of the approaches proposed in \cite{Fazel11}. Therein, a RS procedure, allowing only a randomly chosen sub-set of sensor to acquire the measurement, is coupled with both a  TDMA scheme  and a random  access scheme.

In the RS/deterministic access (RD)  scheme, the FC randomly chooses a set of \(M\) sensors, \(M\) being a sufficient number of measurements for satisfactory field reconstructions and it broadcasts the addresses of eliged nodes through the network. The selected nodes acquire the measurements and   transmit their readings to the FC via a TDMA deterministic access scheme. 

\newcommand{\myCmn}{\lmathcal{C}_{M;N}}
\newcommand{\myKmn}{K_{M;N}}
As in this scheme only \(M\) nodes need to share the TDMA frame, the packet transmission time is \(\myTprd=\myTc/M\). Consequently, the occupied bandwidth for the RD scheme is
\begin{equation}
\myBrd=M\dfrac{L}{\myTc}
\label{eq:myBrd}
\end{equation}
In order to evaluate the consumed energy, let us consider the  set \(\myCmn\) collecting all  the possible configurations of \(M\) out of \(\myN\)  nodes.
The energy consumption of a given configuration \(c \in \myCmn\) of \(M\) is expressed as
\[
E^{(c)}=\sum_{k_c}\myG \;\mydap{k_c}^2\myTprd
\]
where the sum over the index \(k_c\) spans  the \(M\) sensors within the combination \(c\).
Thereby,  the energy consumption of each combination    depends on the distances of the \(M\) nodes from the FC.
In this respective, we evaluate the  energy consumption of the RD scheme as the average over all the possible combinations: 
\begin{equation}
\myErd=\dfrac{1}{\myKmn}\sum_{c\in \myCmn}E^{(c)}
\label{eq:Erdfirst}
\end{equation}
where \(\myKmn={\myN\choose M}\) is the  cardinality of \(\myCmn\).  
After  rewriting the above equation as
\begin{equation*}
\myErd=\dfrac{1}{\myKmn}\sum_{c\in \myCmn}\sum_{k_c}\myG \;\mydap{k_c}^2\myTprd
\end{equation*}
we recognize that in the overall sum over the \(\myKmn\) combinations, the energy spent by each and every network sensor appears in   \({\myN-1\choose M-1}\) terms, corresponding to the  combinations  it belongs to.
Therefore, the above sum can be rewritten as
\begin{equation}
\myErd=\dfrac{1}{\myKmn}{\myN-1\choose M-1}\sum_{k=1}^{N}\myG \;\mydap{k}^2\myTprd
\label{eq:Erdsecond}
\end{equation}
Following the same steps already detailed for the conventional network we have
\begin{equation}
\myErd=\dfrac{M}{\myN}\dfrac{\alpha_1 \alpha_2 (\alpha_1^2 +\alpha_2^2 )}{48}\myN^2\myTprd
\label{eq:Erdthird}
\end{equation}
which, given that \(\myTprd=\myTc/M\), reads as:
\begin{equation}
\myErd=\dfrac{\alpha_1 \alpha_2 (\alpha_1^2 +\alpha_2^2 )}{48}\myN\myTc
\label{eq:Erdfourth}
\end{equation}


The energy consumption and occupied bandwidth performance of the  scheme in \cite{Fazel11} need to be addressed in a slightly different way \wrt the previous cases.
CS theoretic results determine the number \(M\) of measurements needed at the FC to correctly restore the sensed field; within an observation time, this constraint corresponds to a  required percentage \(\myqs\)  of correctly received samples  at the FC.
Because of possible collisions, the required percentage \(\myqs\)  needed at the FC does not translate straightforwardly into a sensing probability \(\myps\).

In \cite{Fazel11}, the authors establish the following relationship among the sensing probability \(\myps\) and the probability \(\myqs\) of correct packet reception:
\begin{equation}
\myqs
=\myps\exp\left(-\myps\dfrac{2\myL\myN}{\myTc B - \myL}\right)
\label{eq:qs}
\end{equation}
being \(B\) the available bandwidth. Remarkably, given \(\myqs\), a minimum bandwidth is required in order to assure that  a feasible value of \(\myps\) exists. 
Thereby,   if the available bandwidth is not accurately dimensioned,  small values of \(\myps\) do not provide enough measurements at the FC, whereas large values of \(\myps\)  cause too many collisions.

\newcommand{\myBmin}{\myBrr^\text{min}}
In \cite{Fazel11}, the authors provide  
an expression,  standing for small values of \(\myqs\), of the minimum  bandwidth as a function of the desired \(\myqs\).
Following the guidelines in \cite{Fazel11}, we have extended such result to accommodate  for large values of \(\myqs\), too.
Specifically, we came up with the  relation:
\begin{equation}
\myBmin=\begin{cases}
\dfrac{\myL}{\myTc}\left(2\myN e \cdot \myqs + 1\right)\;\;\;&\text{if}\;\;\;\myqs\le e^{-1}\\[1em]
\dfrac{\myL}{\myTc}\left(\dfrac{2\myN}{-\ln{\myqs}} + 1\right)\;\;\;&\text{if}\;\;\;\myqs\ge e^{-1}\\
\end{cases}
\label{eq:Bmin}
\end{equation}

The RS settings are therefore assigned as follows. Firstly, the desired \(\myqs\) is fixed according to the reconstruction quality constraints; secondly, the minimum needed bandwidth is evaluated according to \eqref{eq:Bmin}; finally, the selected bandwidth  value is employed  to derive the needed sensing probability \(\myps\) using \eqref{eq:qs}.

With these positions,  the packet transmission time \(\myTprr\) is determined from the employed bandwidth as \(\myTprr\myL/\myBmin\).
Besides,  the energy consumption is determined by the value of \(\myps\); the 
 average network consumed energy 
is evaluated as
\begin{equation}
\myErr=\text{E}\left\{\sum_k \myps \myPk\myTprr \right\}=\dfrac{\myps}{24}\myN^2\myTprr
\label{eq:enrr2}
\end{equation}

By explicating the value of \(\myTprr\), the expression in \eqref{eq:enrr2}, depending on the value of \(\myqs\), reads as follows:
\begin{equation}
\myErr=\begin{cases}
\dfrac{\myps \myN^2}{24}\dfrac{\myTc}{\left(2\myN e \myqs + 1\right)}\;\;\;&\text{if}\;\;\;\myqs\le e^{-1}\\[1em]
\dfrac{\myps \myN^2}{24}\dfrac{\myTc}{\left(\dfrac{2\myN}{-\ln{\myqs}} + 1\right)}\;\;\;&\text{if}\;\;\;\myqs\ge e^{-1}\\
\end{cases}
\label{eq:energyRandom}
\end{equation}
For high values of \(\myN\), the expressions in \eqref{eq:energyRandom}, can be globally approximated as
\begin{equation}
\myErr\simeq\dfrac{\myps}{48} \myN\myTc
\label{eq:enrr3}
\end{equation}

\subsection{Further remarks}
Before turning to the numerical performance evaluation, a few remarks are in order.

The above analysis has pointed out that 
the consumed energy and allocated bandwidth adopting the Radon-like CS scheme grow only with the square root \(\sqrt{N}\) of the network size, where as those of selected state-of-the-art approaches vary with the power \(N\).  
Therefore, the Radon-like CS scheme 
 yields a reduced energy consumption for each node, as well as a parsimonious bandwidth use
for collecting data over all the grid. The gain is more and more evident as the network size (\ie the covered area) increases.

The actual gain in terms of energy and bandwidth depends on the constants \(\mygammap,\mydeltap\) which grow with the number of considered projections. 
The advantages of the Radon-like CS scheme are expected to be  evident on spatially sparse signals, where low values of \(P\) (\eg \(P=3\)) enable reconstruction, whereas RS data gathering algorithm  require a high percentage of the samples to reach the FC in order to achieve satisfying reconstruction results \cite{Luo10}. 

As far as the medium access scheme are concerned, the herein devised Radon-like CS scheme is realized via a TDMA technique, just as the RD scheme; therefore, it implies an effort of synchronization and scheduling. Nonetheless, different data gathering procedures can be envisaged realizing the Radon-like CS using a random access criterion; this issue is left for further studies.

Finally, these results depend on the peculiar structure of the Radon-like sensing matrix \(\Phi_R\) and although derived for a particular data gathering algorithm, can be generalized to different Radon-like CS measurements computation schemes.

%


\section{Numerical simulations}
\label{sec:numres}
In this Section, we analyze the performance of our Radon-like CS scheme both in terms of reconstruction accuracy and of employed network resources. Firstly, in Subsect.\ref{subsect:RecAcc} we present a few results showing that an accurate reconstruction of a spatially sparse signal can be achieved by the \RCSM using a feasible number of Radon-like projections. Secondly, in  Subsect.\ref{subsect:Eff} we demonstrate the energy and bandwidth gain achievable when adopting the Radon-like CS scheme in a WSN.

\subsection{Radon-like CS reconstruction accuracy}
\label{subsect:RecAcc}
With reference to the network model in Sect.\ref{sect:model}, we consider a WSN made up by a square grid of  \(\myN=64\times64=4096\) sensors.
The sensed  field \(z[n_1,n_2]\) is built up by 7 repetitions of an elementary \(5\times 5\) pattern, differently scaled by factors   in the range \((0.5-0.85]\);
Fig.\ref{fig:orfield}-a) describes an example of the field \(z[n_1,n_2]\).
This field adhere to the so called pulse stream signal model described in \cite{bar11}, and the therein presented Algorithm 2 is employed for computing the reconstructed field \(\hat{z}[n_1,n_2]\).

We have first evaluated the reconstruction accuracy of the  Radon Like CS scheme, using  different number \(P\) of projections, under the assumption of noise free observations.
Specifically, we have tested the reconstruction accuracy when only projections along rows and columns of the network grid are considered \((\myP=2, M=128)\), when  projections along the rows, the columns and along the \(\pi/4\) oriented diagonal of the grid are considered \((\myP=3, M=255)\) and finally when also projections along  the \(3\pi/4\) oriented diagonal are considered \((\myP=4, M=382)\).

Tab.\ref{fig:accuracy64} reports the MSE of the reconstructed field \(\vetzh\), \ie \(MSE=(\vetz-\vetzh)\tras (\vetz-\vetzh)\), achieved after 20 iterations of the reconstruction Algorithm 2 and averaged over 10  runs corresponding to different pulse locations. Results in Tab.\ref{fig:accuracy64} show the effectiveness of the Radon Like CS scheme in sensing a sparse field with a restrained number of measurements.
\mytab{htb!}
\centering
\begin{tabular}{|c|c|c|c|}
\hline
\( M\)&\(128\) & \(255\)& \(382\)\\
\hline
\(P\)        &2 & 3& 4\\
\hline
\(\vartheta_p\)        &\(0, \pi/2\) & \(\vartheta_p=0, \pi/2, \pi/4\)& \(\vartheta_p=0, \pi/2, \pm\pi/4\)\\
\hline
Radon Like CS&0.0056&0.0019&0.0011\\
\hline
RS&13.3&7.604&4.721\\
\hline
\end{tabular}
\caption{Reconstruction accuracy (MSE) obtained by the Radon-Like CS scheme and by the RS scheme (\(\myN=64\times64=4096\)).}
\label{fig:accuracy64}
\mytabend

\par\noindent
To visually assess the Radon Like CS scheme reconstruction accuracy, we show in Fig.\ref{fig:orfield}-b) the reconstructed field \(\hat{z}[n_1,n_2]\) obtained by measuring the field in Fig.\ref{fig:orfield}-a) with \(\myP=3\) projections.

For the sake of comparison we have also evaluated the reconstruction accuracy obtained by the RS \cite{Fazel11} scheme under the same experimental settings. In Tab.\ref{fig:accuracy64}  we recognize that, for the selected range of measurements \((M\leq382)\) the RS doesn't allow to capture the sparse nature of the sensed field. To obtain the same reconstruction accuracy of the Radon Like CS scheme, namely a \MSE equal to \(0.0036\), the RS requires be run with a number of measurements \(M\simeq2500\approx 50\%N\).

\protect\myfig{htb!}
\centering
\begin{tabular}{cc}
 \gseteps{0cm}{4cm}{4cm}{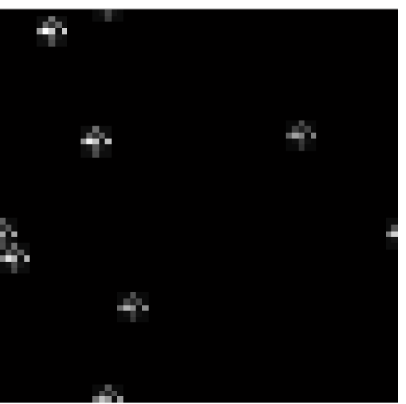} &
 \gseteps{0cm}{4cm}{4cm}{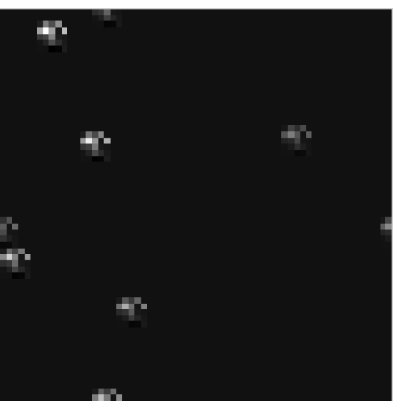}\\
a)&b)
\end{tabular}
\protect\caption{Original field \(z[n_1,n_2]\) (a) and field \(\hat{z}[n_1,n_2]\)(b) after sensing performed according to  the Radon Like CS with \(\myP=3\) projections.}
\vspace*{-0.5em}
\label{fig:orfield}
\myfigend

%
%
%
%
Similar results have been obtained by scaling the number of network nodes to \(\myN=80\times80=6400\) and considering a sensed field \(z[n_1,n_2]\) composed by 8 repetitions of the elementary pulse. The \MSE of the Radon-like CS scheme, averaged over 10 runs, is reported in Tab.\ref{fig:accuracy80}.
Also in this case the MSE obtained by the Radon Like CS scheme with \(\myP=\{2, 3, 4\}\), corresponding to \(M=\{160, 319, 478\}\) measurements, proves to exhibit satisfactory reconstruction quality. For comparison sake, we observe that in these experiments, \(M=3500\) were needed by the RS to achieve analogous performance, namely a \MSE equal to \(0.006231\).
\mytab{htb!}
\centering
\begin{tabular}{|c|c|c|c|}
\hline
\(M\) &\(160\) & \(319\)& \(478\)\\
\hline
\(P\)        &2 & 3& 4\\
\hline
\(\vartheta_p\)        &\(0, \pi/2\) & \(\vartheta_p=0, \pi/2, \pi/4\)& \(\vartheta_p=0, \pi/2, \pm\pi/4\)\\
\hline
Radon Like CS&0.0045&0.00152&0.000901\\
\hline
RS&8.602&6.472&4.506\\
\hline
\end{tabular}
\caption{Reconstruction accuracy (MSE) obtained by the Radon Like CS scheme and by the RS scheme (\(\myN=80\times80=6400\)).}
\label{fig:accuracy80}
\mytabend

\par\noindent
Finally, we have tested the Radon Like reconstruction accuracy when noisy acquisition are considered so that the CS measurements can be modeled as:
\[
\vety=\myPhi\vetz+\vet{n}
\]
where the vector \(\vet{n}\) gathers samples of white zero mean Gaussian noise with variance \(\sigma^2_n\).
In Tab.\ref{tab:noisy} we report the reconstruction accuracy obtained when acquiring with \(\myN=64\times64=4096\) sensors a field composed by 7 patterns via the Radon Like CS scheme for different values of \(\sigma^2_n\).
\mytab{htb!}
\centering
\begin{tabular}{|c|c|c|c|}
\hline
\(M\) &\(128\) & \(255\)& \(382\)\\
\hline
\(P\)        &2 & 3& 4\\
\hline
\(\vartheta_p\)        &\(0, \pi/2\) & \(\vartheta_p=0, \pi/2, \pi/4\)& \(\vartheta_p=0, \pi/2, \pm\pi/4\)\\
\hline
 \(\sigma^2_n=0.5\)&0.0065&0.0034&0.0033\\
\hline
 \(\sigma^2_n=0.7\)&0.007&0.0039&0.0036\\
\hline
\end{tabular}
\caption{Reconstruction accuracy (MSE) obtained by the Radon Like CS;  noisy measurements (\(\myN=80\times80=6400\)).}
\label{tab:noisy}
\mytabend
Results in Tab.\ref{tab:noisy} show how the presence of noise in the acquisition process  does not severely affect the reconstruction performance of the Radon Like CS approach.

To recap, the above results show that,
as a rule of thumb, Radon Like CS requires 
\(M\approx \myP \sqrt{N} \) measurements for representing a spatially sparse field, whereas the RS requires a large percentage of the measurements \(M_{RS} \approx \alpha N\) to be correctly received (\eg, \(P=3\) and \(\alpha=50\%\) in the above experiments).
Overall, the Radon Like CS scheme allows sensing and reconstructing of a spatially sparse field with far less measurement \wrt to state of the art techniques such as the RS presented in \cite{Fazel11}. 


\subsection{Radon Like CS efficiency}
\label{subsect:Eff}

We now show that, besides using a restrained number of  measurements, the Radon Like CS presents significant advantages in terms of energy and bandwidth needed to disseminate sensors readings towards the FC.


We evaluate the performance of Radon Like CS data gathering scheme (RL) by evaluating the consumed energy and occupied bandwidth under the following assumptions
\begin{itemize}
\item[\textit{a1)}] the number of measurements is chosen large enough to yield a satisfactory reconstruction accuracy, quantified by a \(\MSE\leq 10^{-3}\);
\item[\textit{a2)}] the coherence time  of the sensed field is fixed to \(\myTc=2500s\);
\item[\textit{a3)}] the dimension of the transmitted packet is set to \(\myL=1Kb\).
\end{itemize}

With reference to the experimental setting described in Sect.\ref{subsect:RecAcc}, condition a1) implies \(\myP=3\) projections.

We start by computing the occupied bandwidth required by the  RL scheme 
to guarantee correct sensing and reconstruction of a spatially sparse field.
Fig.\ref{fig:bandwidth} plots the occupied bandwidth evaluated according to \eqref{eq:brad} vs the network size in terms of number of sensors. 

For the sake of comparison, in the same figure we also report the performance of the RS, implemented both using a deterministic access scheme (RD in the legend) and the random access scheme (RR in the legend) described in \cite{Fazel11} and analyzed in Sect.\ref{sub:conv}; besides, we report the results  for the conventional network (DD in the legend) described in Sect.\ref{sub:conv}.
The occupied bandwidth is computed  according to  \eqref{eq:Bconv}, \eqref{eq:myBrd} and \eqref{eq:Bmin} for the RL, DD, RD and RR scheme respectively, under the same assumptions a1)-a3)
.
These settings imply \(p_s=0.5\) for the RD scheme and \(q_s=0.5\) for the RR one, whereas the conventional network, bringing all the measurements to the FC, always achieves perfect reconstruction.

\unepsperriga{9cm}{6cm}{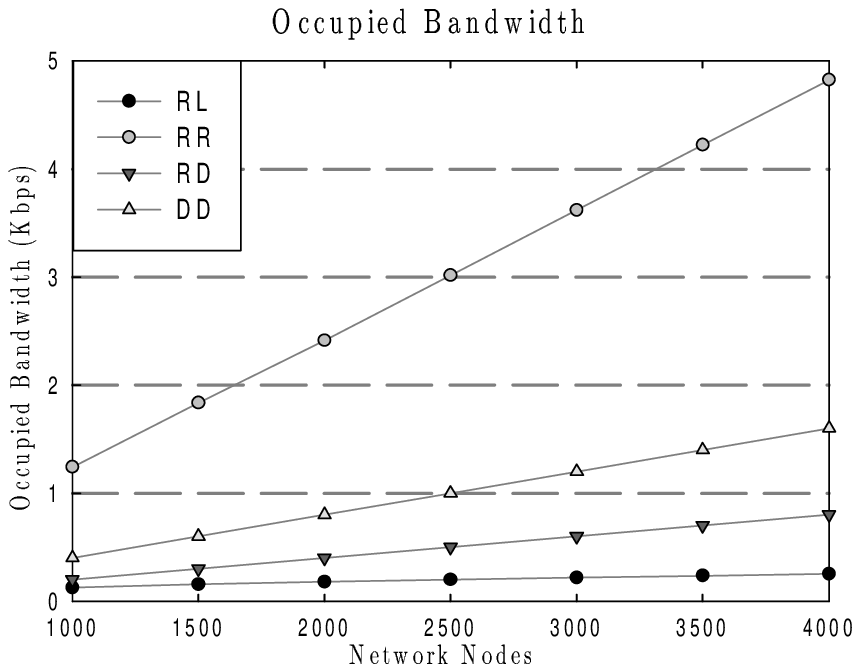}{\vspace{-0em}\caption{Bandwidth occupancy versus number of nodes N. \label{fig:bandwidth}}}
\unepsperriga{9cm}{6cm}{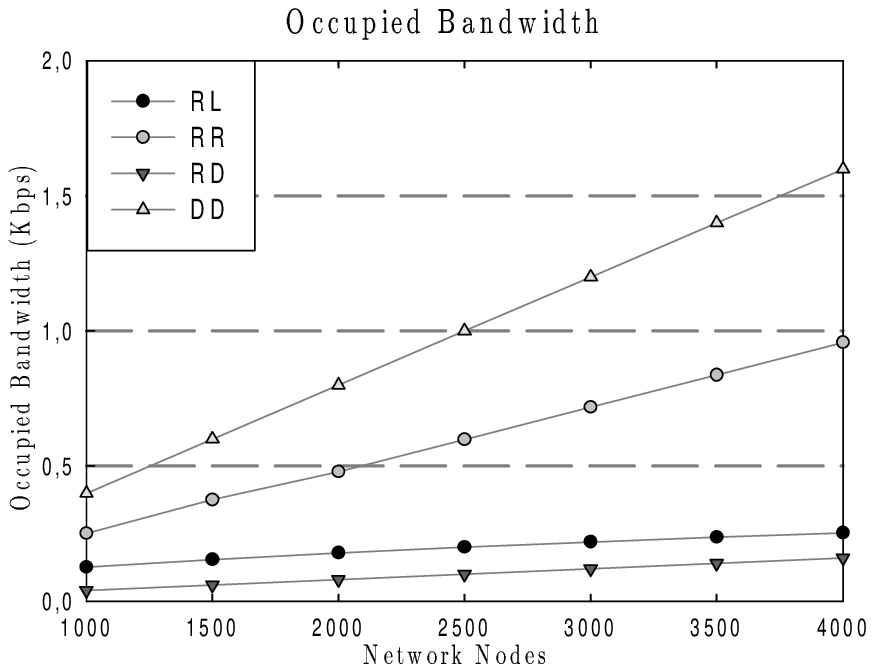}{\vspace{-0em}\caption{Bandwidth occupancy versus number of nodes N. \label{fig:band10}}}
\unepsperriga{9cm}{6cm}{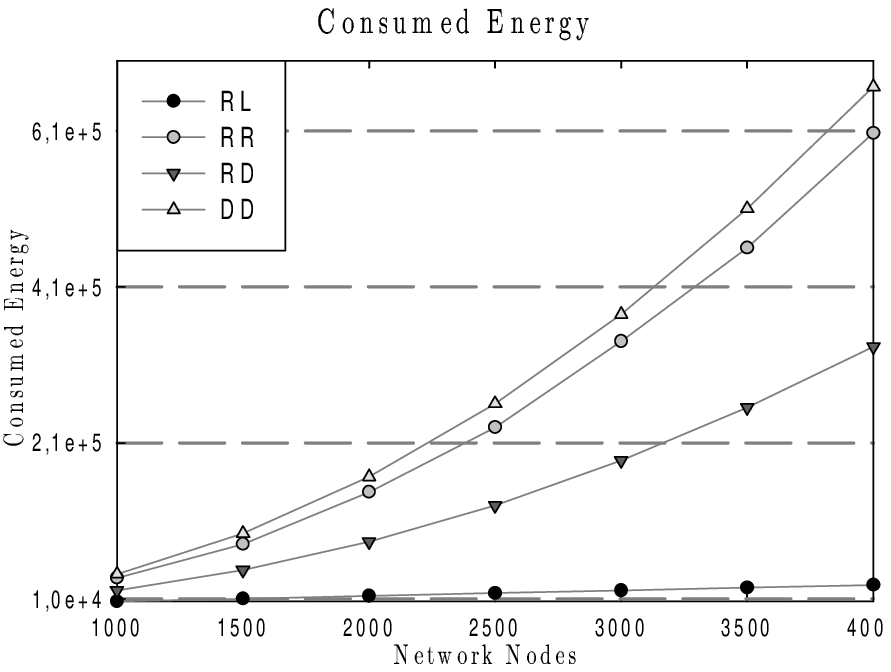}{\vspace{-0em}\caption{Energy consumption versus number of nodes N. \label{fig:energy}}}
\unepsperriga{9cm}{6cm}{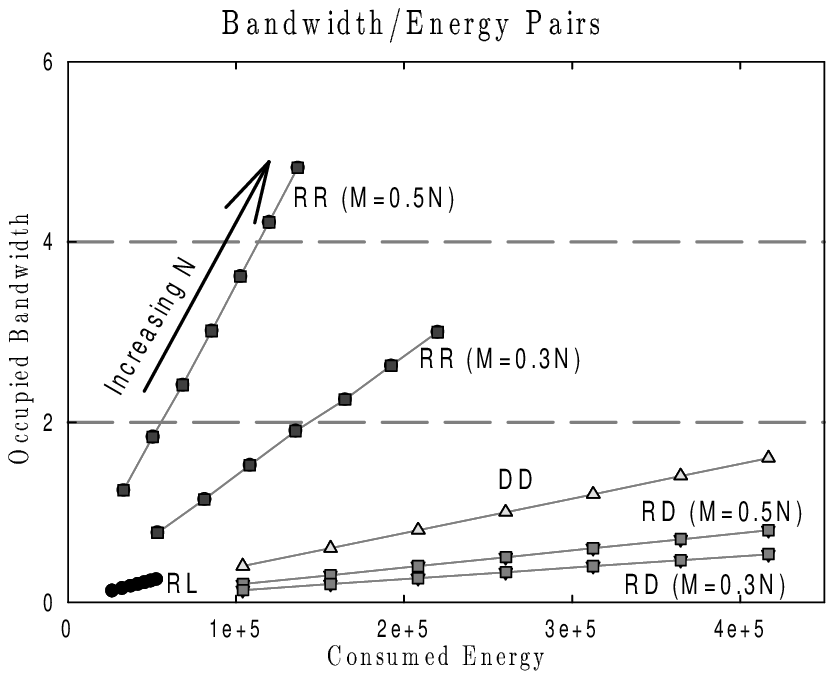}{\vspace{-0em}\caption{Energy/Bandwidth pairs. \label{fig:scatter}}}
The bandwidth employed for the Radon-like data gathering scheme is significantly reduced, not only \wrt the conventional network, but also \wrt the RS, both using deterministic and random access. The above results, obtained by considering different number of measurements and equal reconstruction accuracy for the three algorithms, are explained by the efficiency of the Radon-like data collection algorithm, exploiting only single hop, parallel data transmission. 

We have then pushed further the comparison to find under which conditions the Radon-like and the RS schemes equally perform. In doing so, we have evaluated the  bandwidth occupied by  the RD and RR schemes  when only 10\% of the sensor readings are required to be correctly received by the FC. Fig.\ref{fig:band10} reports these results showing that the RL scheme still favorably compares with the RR scheme, while requiring just slightly higher bandwidth \wrt the RD scheme. Therefore, the RL scheme overcomes the RD and RR ones unless they use a very low percentage of the sensors' readings for field reconstruction. 

Let us now compare the energy consumption 
under the assumption that the occupied bandwidth, and consequently 
 the packet transmission time \(\myTp=\myL/\myB\), are the same for the different schemes.

Let us remark that each scheme requires a different number of transmissions in order to let the FC acquire the needed measurements. Thereby,   the overall process of sensing is  accomplished in different time intervals by the different data gathering techniques. 
In these experiments, we have set \(\myTp=0.61\) so that also for the DD scheme, comprising the maximum number of packet transmission, the  data gathering time meets the constraint on the field coherence time (namely \(\myN T_p\leq T_c\)) even for \(\myN=4000\).
Fig.\ref{fig:energy} reports the energy consumed for different number of network sensors \(\myN\); we recognize that the employment of the RL scheme  drastically reduces the energy consumed by the data gathering algorithm.

Finally, we are interested in comparing the resource employed by the different schemes to accomplish the sensing process exactly in the same time. To reach this goal, each scheme occupies a different occupied bandwidth.
In Fig.\ref{fig:scatter} we illustrate the bandwidth-energy scatterplot  of the RL, RR, RD and DD schemes  for different network sizes.
For the RD and RR we have considered two cases, that is when 50\% and 30\%  of the sensors readings are required at the FC for satisfactory field reconstruction.

The energy/bandwidth pairs draw different trajectories  while the number of nodes \(\myN\) increases. We clearly recognize a systematic energy and bandwidth saving of the RL scheme  \wrt to the competitors.
Moreover, both the consumed energy and the occupied bandwidth exhibit far smaller variations with the number of network nodes than what happens with the RD, RR and DD schemes, making the RL approach fully scalable in terms of network nodes.

As a final remark, we observe that energy and bandwidth gain yielded by the Radon-like approach is directly based upon 
the favorable matching between the Radon-like matrix structure and the
spatially sparse structure of the sensed field.
Based on these results, we envisage a twofold extension of our work, that is 
\begin{itemize}
\item in accounting an irregular sampling grid; 
\item in exploiting non straight  sampling path.
\end{itemize} 
These conditions may be encountered, for instance, in  vehicular networks or citizen sensing networks \cite{Kan11}, where the disposition of the nodes are far from being regular, and the sampling path should adapt to the routing paths, which  in turn basically depend on the streets and buildings  disposition.  

 To sum up, we consider this contribution as an intermediate step towards the finding of a general relation between the compressive sensing  of a finite innovation rate signal and its realization by efficient routing algorithms in a realistic  WSN scenario.

\section{Conclusion}\label{sec:concl}
In this paper, we have addressed the efficient compressive sampling  of spatially sparse signals in sensor networks.
Specifically, we have introduced a peculiar CS sampling scheme for spatially sparse bidimensional signals. We have analytically demonstrated that our scheme satisfies the theoretical conditions required for CS signal reconstruction. Then, after devising a distributed data gathering scheme for  collecting the CS measurements in a WSN, we have characterized the scheme both in terms of consumed transmission energy and occupied bandwidth. The scheme outperforms state-of-the-art schemes for spatially sparse fields, and it  represents an intermediate step towards the definition of routing procedures well-suited to the characteristics of the signal a realistic sensor network is faced to.

\renewcommand{\theequation}{\Roman{section}.\arabic{equation}}
\renewcommand{\thesection}{Appendix \Roman{section}}
\renewcommand{\thesection}{Appendix \Alph{section}}
\setcounter{section}{0}
\setcounter{subsection}{0}
\setcounter{equation}{0}

\section{RIP property of the Radon-like measurements matrix.}
\label{app:proof}

\textbf{Property:} If   the nonzero  entries in \(\Phi_R\) are i.i.d. zero mean Gaussian random variables with equal variance \(\sigma^2_{\varphi}=1/P\), the following  concentration inequality stands:
\begin{equation}
\label{eq:appci}
\Pr\{|\myEy-\myEz|\geq\delta\}\leq \epsilon
\end{equation}
provided that \(P\geq 2 K^2 C_2^2 \log(2/\epsilon)/\delta^2 \) being \(C_2\) a suitable constant.

\newcommand{\dimgae}{1}
\textbf{Demonstration:} 
\ifthenelse{\dimgae=0}{
The condition in \eqref{eq:appci} 
can be demonstrated as follows. Let us consider the sample energy \(\myEy\) of the measurements. Being \(\myEy\) a sample moment, we here invoke its asymptotic  normal distributions  \cite{Kendall}. %
Although this hypothesis is not necessary for the RIP to stand, it allows us to straightforwardly evaluate the minimal number of projections \(P\) required for \(K\)-sparse field reconstruction, and therefore we retain it in the followings. 

The Chernov bound \cite{proakis} for a normal random variable  establishes that  the probability that the random variable differs from its  mean  is limited by a  term exponentially decaying with its variance. 
By application of the Chernov bound to the random variate \(\myEy\), we obtain:
 \[\Pr\{|\myEy-{\E{\myEy}}|\geq\delta\}\leq 2\cdot\exp^{-\delta^2/2\sigma^2_{\myEy}}\]
Thereby, in order to demonstrate \eqref{eq:ci} it suffices  to show that \(\E{\myEy}=\myEz\) and \(\gVar{\myEy}=o(P)\) for the case under concern.

By definition we have
\[
\E{\myEy}
=\E{\sum_{p=0}^{P-1}
\sum_{m=0}^{K{(p)}-1} \left(\sum_{i} \varphi^{(p)}_{m}[i] z^{(\vartheta_p)}[i,m]\right)^2 }
\]

Since  \(\varphi^{(p)}_{m}[i]\)  are i.i.d. zero mean random variables, the cross terms  \(\varphi^{(p)}_{m}[i]\cdot  \varphi^{(p)}_{m}[j]\) contributing to \(\myEy\) average to zero and we obtain  
\[
\E{\myEy}
=\sum_{p=0}^{P-1}
\sum_{m=0}^{K{(p)}-1}
\sum_{i} \E{\varphi^{(p)}_{m}[i]^2 } z^{(\vartheta_p)}[i,m]^2
\]
Observing that 
\[\sum_{p=0}^{P-1}
\sum_{m=0}^{K{(p)}-1}
\sum_{i}
z^{(\vartheta_p)}[i,m]^2=\sum_{p=0}^{P-1} \cdot \myEzthp=P \cdot \myEz\] 
under the hypothesis \(\sigma^2_{\varphi}=1/P\) it results
\[
\E{\myEy}=P\cdot
\sigma_{\varphi}^2\cdot\myEzthp=\myEz.
\]
As far as the evaluation of the variance \(\gVar{\myEy}=\E{\myEy^2}-\E{\myEy}^2\) is concerned, we evaluate the second order moment
\[\E{\myEy^2}=\E{\sum_{m_1=0}^{M-1}\sum_{m_2=0}^{M-1}y_{m_1}^2 y_{m_2}^2}.\]
Recalling that \(y^{(\vartheta_p)}[m]=\sum_{i} \varphi^{(p)}_{m}[i] z^{(\vartheta_p)}[i,m]\), and that
\[\E{\varphi^{(p)}_{m}[i]\varphi^{(p)}_{m}[j]}= \sigma_{\varphi}^2 \delta_{i-j}\] being \(\delta_{i-j}\) the Kronecker delta,
after some  algebra we obtain%
\footnote{ We compactly denote the nested sums \(\sum_{i_1}\dotsc\sum_{j_2}\) sharing the same bounds \(0,\myJ-1\) by \(\sum_{i_1,\dotsc j_2}\).}
\newcommand{\stfz}[1]{z_{#1}}

\begin{equation}\begin{split}
&\E{\myEy^2}=\sum_{p=0}^{P-1}\sum_{m_1=0}^{K{(p)}-1}\sum_{m_2=0}^{K{(p)}-1}
\sum_{i_1,j_1,i_2,j_2} 
\\&\phantom{\sum}z^{(\vartheta_{p})}[i_1,m_1]z^{(\vartheta_{p})}[j_1,m_1]z^{(\vartheta_{p})}[i_2,m_2]z^{(\vartheta_{p})}[j_2,m_2])\cdot
\\&\phantom{\sum}
\E{\varphi^{(p)}_{m_1}[i_1]\varphi^{(p)}_{m_1}[j_1]  \varphi^{(p)}_{m_2}[i_2]\varphi^{(p)}_{m_2}[j_2] }
\\&+\underbrace{\sum_{p_1=0}^{P-1}\sum_{p_2=0}^{P-1}}_{p_1\neq p_2}\sum_{m_1=0}^{K{(p_1)}-1}\sum_{m_2=0}^{K{(p_2)}-1}
\sum_{i_1,i_2}
\\&\phantom{\sum}(z^{(\vartheta_{p_1})}[i_1,m_1])^2 (z^{(\vartheta_{p_2})}[i_2,m_2])^2 \cdot \sigma_{\varphi}^4
\end{split}
\end{equation}
\par\noindent
To proceed, let us consider the fourth order moment 
\[
\E{\varphi^{(p)}_{m_1}[i_1]\varphi^{(p)}_{m_1}[j_1]  \varphi^{(p)}_{m_2}[i_2]\varphi^{(p)}_{m_2}[j_2] }\] 
of the random entries of the \(p\)-th row of \(\Phi_R\).
Since all the entries of \(\Phi_R\) are zero mean i.i.d. random variables, the above moment takes only two non zero values, namely  
\(\mu_{\varphi}^{(4)}=\E{(\varphi^{(p)}_{m}[i])^4 }\) when  all the four  variates coincide, and
\(\sigma_{\varphi}^4\) for pairwise coincident random variables.

Based on these observations, the above expected value can be rewritten as
\begin{equation}\begin{split}
&\E{\myEy^2}=\sum_{p=0}^{P-1}\sum_{m_1=0}^{K{(p)}-1}\sum_{m_2=0}^{K{(p)}-1}
\sum_{i_1,i_2}  
\\&\phantom{\sum}(z^{(\vartheta_{p})}[i_1,m_1])^2 (z^{(\vartheta_{p})}[i_2,m_2])^2\cdot
\\&\phantom{\sum}
\left(\E{(\varphi^{(p)}_{m_1}[i_1])^2  (\varphi^{(p)}_{m_2}[i_2])^2 }\pm \sigma_{\varphi}^4 \delta_{i_1-i_2} \delta_{m_1-m_2}\right)
\\&+\underbrace{\sum_{p_1=0}^{P-1}\sum_{p_2=0}^{P-1}}_{p_1\neq p_2}\sum_{m_1=0}^{K{(p_1)}-1}\sum_{m_2=0}^{K{(p_2)}-1}
\sum_{i_1,i_2}
\\&\phantom{\sum}(z^{(\vartheta_{p_1})}[i_1,m_1])^2 (z^{(\vartheta_{p_2})}[i_2,m_2])^2 \cdot \sigma_{\varphi}^4
\\&= \left(\mu_\varphi^{(4)}- \sigma_{\varphi}^4\right) \cdot \sum_{p=0}^{P-1}\sum_{m=0}^{K{(p)}-1}\sum_{i}  
(z^{(\vartheta_{p})}[i,m])^4 
\\&+
\left(
\sigma_{\varphi}^2 \sum_{p=0}^{P-1}\sum_{m=0}^{K{(p_1)}-1}\sum_{i}
(z^{(\vartheta_{p})}[i,m])^2  
\right)^2
\end{split}
\end{equation}

Finally, for the variance \(\gVar{\myEy}\) we have
\begin{equation}
\gVar{\myEy}= \left(\mu_\varphi^{(4)}\!-\! \sigma_{\varphi}^4\right) \cdot \sum_{p=0}^{P-1}\sum_{m=0}^{K{(p)}-1}\sum_{i}  
\left(z^{(\vartheta_{p})}[i,m]\right)^4 
\label{eq:gvar}
\end{equation}

Recalling that \(\sigma_{\varphi}^2=1/P\), and that \(\mu_\varphi^{(4)}=3 \sigma_{\varphi}^4\;\) due to the normal distribution
%
 of the entries of \(\Phi_R\), we obtain:
\begin{equation}
\gVar{\myEy}= \dfrac{1}{P^2} \sum_{p=0}^{P-1}\sum_{m=0}^{K{(p)}-1}\sum_{i}  
\left(z^{(\vartheta_{p})}[i,m]\right)^4 
\label{eq:gvarexplicit}
\end{equation}
Since \( \myquad{z}{n_1}{n_2}^4\) consists of \(K\) non zero samples, the two inner sums in \eqref{eq:gvarexplicit} span over \(K\) terms as a maximum, and the variance \(\gVar{\myEy}\) is upper bounded by:
\begin{equation}
\gVar{\myEy} \leq\dfrac{1}{P} K \max_{(n_1,n_2)} \myquad{z}{n_1}{n_2}^4=\dfrac{1}{P} K C_4 
\label{eq:gvarfinal}
\end{equation}
In \eqref{eq:gvarfinal} we recognize that the variance \(\gVar{\myEy}\) decays as \(1/P\). 
Furthermore, by comparison of \eqref{eq:gvarfinal} and \eqref{eq:ci} we recognize that the RIP is verified
provided that the number of projections \(P\) satisfies:
\[P\geq 2 K C_4 \log(2/\epsilon)/\delta^2 \]
}
{

The condition in \eqref{eq:appci} 
can be demonstrated as follows. Let us consider the sample energy \(\myEy\) of the measurements. Being \(\myEy\) a sample moment, we here invoke its asymptotical normal distributions  \cite{Kendall}. %
Although this hypothesis is not necessary for the RIP to stand, it allows us to straightforwardly evaluate the minimal number of projections \(P\) required for \(K\)-sparse field reconstruction, and therefore we retain it in the followings. 

The Chernov bound \cite{proakis} for a normal random variable  establishes that  the probability that the random variable differs from its  mean  is limited by a  term exponentially decaying with its variance. 
By application of the Chernov bound to the random variate \(\myEy\), we obtain:
 \[\Pr\{|\myEy-{\E{\myEy}}|\geq\delta\}\leq 2\cdot\exp^{-\delta^2/2\sigma^2_{\myEy}}\]
Thereby, in order to demonstrate \eqref{eq:ci} it suffices  to show that \(\E{\myEy}=\myEz\) and \(\gVar{\myEy}=o(P)\) for the case under concern.

By definition we have
\[
\myEy
=\sum_{p=0}^{P-1}\sum_{m=0}^{K{(p)}-1} \left(\sum_{i} \varphi^{(p)}_{m}[i] z^{(\vartheta_p)}[i,m]\right)^2 
\]
Let us denote by
\newcommand{\myythetamp}{y^{(\vartheta_p)}_{m}}
\[
\myythetamp
\ggdef \sum_{i} \varphi^{(p)}_{m}[i] z^{(\vartheta_p)}[i,m], p=0\dots P-1,\; m=0,\dots K{(p)}-1
\]
We recognize that \(\myythetamp\) are 
i.i.d. normal random variables, with zero mean and variance equal to
\[\Var{\myythetamp}=\sigma^2_{\varphi} \sum_{i} \left(z^{(\vartheta_p)}[i,m]\right)^2\]
With these positions, we recognize that 
\[\begin{split}
\E{\myEy}
&=
\sum_{p=0}^{P-1}
\sum_{m=0}^{K{(p)}-1}
\Var{\myythetamp}\\
&=\sigma^2_{\varphi}
\sum_{p=0}^{P-1}\sum_{i} \left(z^{(\vartheta_p)}[i,m]\right)^2
\end{split}
\]
As far as the variance of \(\myEy\) is concerned, we obtain
\[\begin{split}
\Var{\myEy}
&=
\sum_{p=0}^{P-1}
\sum_{m=0}^{K{(p)}-1}
\Var{\left(\myythetamp\right)^2}\\
&=\sum_{p=0}^{P-1}
\sum_{m=0}^{K{(p)}-1}
\left(
\E{\left(\myythetamp\right)^4}-\E{\left(\myythetamp\right)^2}^2
\right)
\end{split}
\]
Since the variables \(\myythetamp\) are zero-mean normally distributed, their fourth-order moments satisfy
\[\E{\left(\myythetamp\right)^4}=3\cdot\left(\Var{\myythetamp}\right)^2\]
so that we obtain
\[
\Var{\myEy}
=2
\sum_{p=0}^{P-1}
\sum_{m=0}^{K{(p)}-1}
\left(
\Var{\myythetamp}
\right)^2
\]
Observing that, for a \(K\)-sparse signal, 
the maximum value of \(\sum_{i_2}\left(
z^{(\vartheta_p)}
[i_2,m]\right)^2\) 
is achieved in case of \(K\) aligned pulses, we recognize that 
the following inequality stands:
\[
\begin{split}
&\left(
\Var{\myythetamp}
\right)^2
=\sigma^4_{\varphi}
\sum_{p=0}^{P-1}
\sum_{m=0}^{K{(p)}-1}
\sum_{i_1} 
\left(
z^{(\vartheta_p)}
[i_1,m]\right)^2
\\&
\phantom{aaaaaaaa}
\cdot
\sum_{i_2}\left(
z^{(\vartheta_p)}
[i_2,m]\right)^2
\\&
\phantom{aaaaaaaa}
\leq \sigma^4_{\varphi}
\sum_{p=0}^{P-1}
\sum_{m=0}^{K{(p)}-1}
\sum_{i_1} 
\left(
z^{(\vartheta_p)}
[i_1,m]\right)^2
\cdot K C_2
\\&
\phantom{aaaaaaaa}
=\sigma^4_{\varphi} P \cdot \myEz \cdot K C_2
\end{split}
\]
being \(C_2=\max_{(n_1,n_2)} \myquad{z}{n_1}{n_2}^2\).

Finally, we  obtain that
the variance \(\gVar{\myEy}\) is upper bounded by:
\begin{equation}
\gVar{\myEy} \leq \dfrac{1}{P}  \cdot K^2 C_2^2
\label{eq:gvarfinal}
\end{equation}
In \eqref{eq:gvarfinal} we recognize that the variance \(\gVar{\myEy}\) decays as \(1/P\). 
Furthermore, by comparison of \eqref{eq:gvarfinal} and \eqref{eq:ci} we recognize that the RIP is verified
provided that the number of projections \(P\) satisfies:
\[P\geq 2 K^2 C_2^2 \log(2/\epsilon)/\delta^2 \]
}


\section{Within WSN Radon-like projections' computation}
\label{app:dga}
Let us  consider a regular network composed by \(\myN=\myI\myJ\) sensors as in Fig.\ref{fig:net} where, without loss of generality, we assume that \(\myI\) and \(\myJ\) are odd valued, \ie \(\myI=2\myItilde+1\), \(\myJ=2\myJtilde+1\), so as to identify a central column where the FC is located.

We  discuss a simple sub-optimal procedure  to collect all the projections to the FC using a TDMA  access scheme.

Let us subdivide the network  in 4 quadrants, and let us  consider first the horizontal projections \(\phor[m]\),  obtained by accumulating randomly weighted values along the network rows.
In each quadrant, 
 the data gathering process starts at the outer nodes. 
The external node in each row measures the field, computes the product of the reading with a randomly selected coefficient, encodes this value in a packet of \(\myL\) bits, and transmits it to the neighboring node in the horizontal direction. The neighboring node, once the packet from the outer node has been received, measures the field, multiplies the reading by the random coefficient, and sums it to the value received by the outer node. The overall process continues until the nodes in the central column are reached by the data flow and are then ready to propagate the projection values to the FC.
Once the FC has received the horizontal projection values from the first quadrant, the same operations are serially performed in the remaining 3 quadrants.

The  projection values of each quadrant are therefore computed by  evaluating  partial sums and  propagating them towards the nodes in the central column; then, the projection values are transmitted to the FC via a multihop route along the central column.

Let us now evaluate the number of transmission required for computing the horizontal projections. For collecting the contributions within a network quadrant we need
\begin{itemize}
\item \(\myItilde\) transmission to reach the central column for each one of the \(\myJtilde+1\) rows;
\item \(\sum_{l=0}^{\myItilde}l\) transmissions to propagate the projections value towards the FC along the central column.
\end{itemize}
Accounting for the four quadrants, the overall number of transmissions for horizontal projections evaluation sums up to
\[
N^{(\pi/2)}_{\text{TX}}=\dfrac{\myI-1}{2}\left(2\myJ+\myI+3\right)
\]
By denoting \(\myI\ggdef\alpha_1 \sqrt{N}\), \(\myJ\ggdef \alpha_2\sqrt{N}\), we can write:
\begin{equation}
N^{(\pi/2)}_{\text{TX}}=(\alpha_1\alpha_2 + \dfrac{\alpha_1^2}{2})\myN + (\alpha_1 - \alpha_2)\sqrt(\myN) - 1.5
\label{transHor}
\end{equation}

\par\noindent
Let us now evaluate the number of time slots in which the \(N^{(\pi/2)}_{\text{TX}}\) transmissions can be performed.
Since signaling occurs between adjacent nodes, the propagation of the information on the different rows can be   scheduled in parallel flows, provided that a suitable inter-row delay  of \(\mytaui\) time slots is introduced to prevent interference among neighboring nodes.

Let us sketch out a possible time scheduling for within quadrant transmission, corresponding to the following gathering protocol:
\begin{itemize}
\item  the data gathering starts at \(\myto=0\), on the first row of the quadrant, \ie the one comprising the FC; the outest node transmits its randomly weighted sensed value in the first time slot; in the second time slot, the second node forwards the sum of the received data and its own, randomly weighted, sensed value; similarly, each node updates and sends the received partial sum; thereby,     the FC retrieves the accumulation after \(\mytf{1}=\myItilde\myTp\), being \(\myTp\) the duration of a time slot;
\item on the second row, the transmission begins after \(\mytaui\) time slots to avoid interference with the first row transmission; then, \(\myItilde\) time slots are needed for the partial sum to reach the central column and  one additional time slot is needed to reach the FC; the propagation ends at \(\mytf{2}=(\mytaui+\myItilde+1)\myTp\);
\item on the \(i\)-th row, the transmission begins after \((i\!-\!1)\cdot \mytaui\) time slots  and the propagation ends at \(\mytf{i}=(i\cdot\mytaui+\myJtilde+i)\myTp\);
\item on the (\(\myJtilde+1\))-th row, transmission to  the FC is accomplished at \(\mytf{\myJtilde+1}=(\myJtilde\mytaui+\myItilde+\myJtilde)\myTp\).
\end{itemize}

For the sake of clarity, we  report in Fig.\ref{fig:datatime} a  scheme summarizing the timing of nodes' transmissions when computing the horizontal projections \(\phor[m]\) within a quadrant of the network.
\unepsperriga{9cm}{6cm}{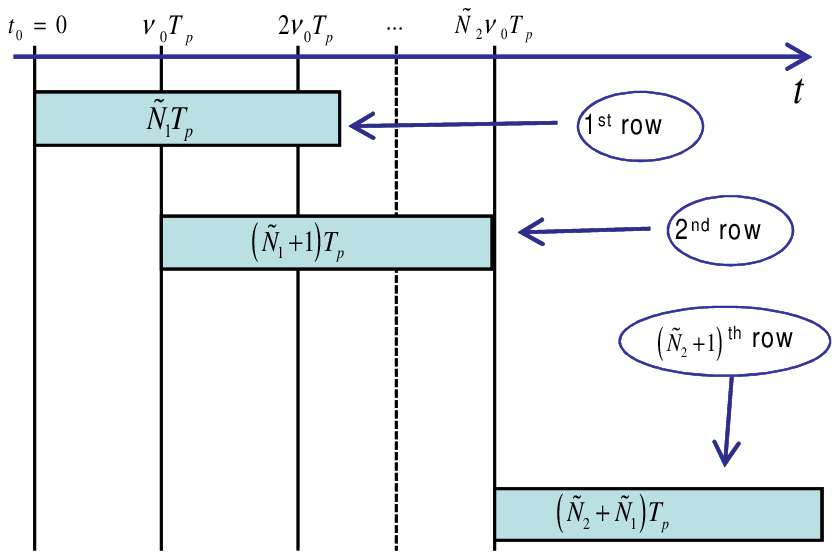}{\caption{Data gathering algorithm: timing of the nodes' transmissions for within wuadrant evaluation of \protect\(\phor[m]\). \label{fig:datatime}}}

The FC is then able to collect all the horizontal projections in a quadrant after \(N_q=\myJtilde\mytaui+\myItilde+\myJtilde\) time slots.
If the quadrants are visited in a serial fashion, the overall number of time slots to compute the horizontal projections accounts for
\[
N^{(\pi/2)}_{\text{TS}}=4(\myJtilde\mytaui+\myItilde+\myJtilde)
\]
Again, by denoting \(\myI\ggdef\alpha_1 \sqrt{N}\), \(\myJ\ggdef \alpha_2\sqrt{N}\), we have:
\begin{equation}
N^{(\pi/2)}_{\text{TS}}=2\sqrt{N}\left[\alpha_2 + \alpha_1(1+\mytaui)\right] -(4+2\mytaui)
\label{slotsHor}
\end{equation}

 The overall protocol for evaluating \(\phor[m]\)  is illustrated in Fig.\ref{fig:datagath}.

An important remark is in order. Despite its simplicity, this basic results highlights one of the major advantages of the Radon-like CS scheme. Since the Radon-like projections can be evaluated by means of information propagation on linear paths in the network,  the number of single-hop transmissions  vary with the product depth and width of the network grid, that is with  the network size \(N\).  On the other hand, since only single-hop transmissions are employed, the transmission can be parallelized and the number of time slots required for computation of an assigned projection set \(y^{(\vartheta_p)}[m]\) vary with the sum of depth and width of the network grid, that is with the square root of the network size \(N\). This intrinsic behavior, that holds for different projections' directions, is the reason why the Radon-like CS scheme will be prove to be both energy and bandwidth efficient.  

With slight modifications, the above described procedure can be extended  to the case of differently tilted paths.
For concreteness sake, we develop in the following the calculations for \(\vartheta_p=0\) (vertical projections), \(\vartheta_p=\pm\pi/4\) (diagonal projections), which have been considered in the simulations described in this paper.

Vertical projections \(\pver[m]\) can be computed using the same protocols adopted for the horizontal projections so that we have:
\[
N^{(0)}_{\text{TX}}=\dfrac{\myJ-1}{2}\left(2\myI+\myJ+3\right)
\]
\[
N^{(0)}_{\text{TS}}=4(\myItilde\mytaui+\myJtilde+\myItilde)
\]
or, equivalently
\begin{equation}
N^{(0)}_{\text{TX}}=(\alpha_2\alpha_1 + \dfrac{\alpha_2^2}{2})\myN + (\alpha_2 - \alpha_1)\sqrt(\myN) - 1.5
\label{transVer}
\end{equation}
\begin{equation}
N^{(0)}_{\text{TS}}=2\sqrt{N}\left[\alpha_1 + \alpha_2(1+\mytaui)\right] -(4+2\mytaui)
\label{slotsVer}
\end{equation}

If only horizontal and vertical projections are considered within the Radon Like CS scheme, stemming from \eqref{transHor} and \eqref{transVer} we can evaluate the overall number of transmissions needed to propagate data towards the FC as
\[
N^{(0, \pi/2)}_{\text{TX}}=\left(2\alpha_1\alpha_2 + \dfrac{\alpha_1^2 + \alpha_2^2}{2}\right)\myN - 3
\]
where we recognize how the required number of transmissions grows linearly with the number of network nodes
\begin{equation*}
\begin{split}
&N^{(0, \pi/2)}_{\text{TX}}\approx\gamma_2\myN\\
&\gamma_2=\left(2\alpha_1\alpha_2 + \dfrac{\alpha_1^2 + \alpha_2^2}{2}\right)
\end{split}
\end{equation*}
For what the number of time slots needed to collect the sensors' readings, referring to  \eqref{transVer} and \eqref{slotsVer}, we have:
\[
N^{(0, \pi/2)}_{\text{TS}}=\left(\alpha_1+\alpha_2\right)\left(4+2\mytaui\right)\sqrt{N} - \left(8+4\mytaui\right)
\]
where we recognize how the required number of time slots grows with the squared root of the network size:
\begin{equation*}
\begin{split}
&N^{(0, \pi/2)}_{\text{TS}}\approx\delta_2\sqrt{N}\\
&\delta_2=\left(\alpha_1+\alpha_2\right)\left(4+2\mytaui\right)
\end{split}
\end{equation*}

\unepsperriga{9cm}{6cm}{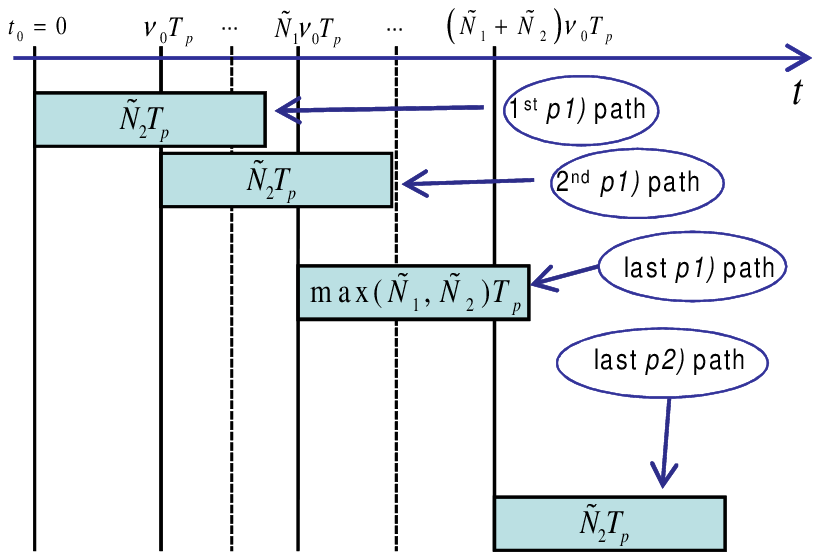}{\caption{Data gathering algorithm: timing of the sensors' transmissions for diagonal projections evaluation. \label{fig:diagtime}}}

For what the diagonal projections \(\pdia[m]\) are concerned, the procedure is similar to the previous case, in the sense that the gathering scheme is aimed at propagating the projection values from the outer row and column of the quadrant to the FC by first reaching the nodes in the inner row and column. We serially perform the accumulations in each quadrant, as described for the horizontal projection, along

\smallskip\par
\ltextit{p1) } all the \(\pi/4\) oriented paths
 that originate from the nodes lying along the outer row of the quadrant and that reach the FC through either the central row or the central column  (solid arrows in Fig.\ref{fig:diag});
\smallskip\par
\ltextit{p2) } all the \(\pi/4\) oriented paths 
 that originate from the nodes lying along the outer column of the quadrant and that reach the FC through either the central row or the central column  (dashed arrows in Fig.\ref{fig:diag}).

The paths  \textit{p1)} -  \textit{p2)} exhibit all a length of at most  \(\max\{\myItilde, \myJtilde\}\).
Each  of the \(\myItilde+\myJtilde+1\) projections are composed by at most  \(\max\{\myItilde, \myJtilde\}\) transmissions, so that the overall number of needed transmissions sums up to
\begin{equation}
\begin{split}
&N^{(\pi/4)}_{\text{TX}}=\left(\myI+\myJ\right)\left(\max\{\myI,\myJ\}-1\right)\\
&\phantom{N^{(\pi/4)}_{\text{TX}}}=(\alpha_1+\alpha_2)\max\{\alpha_1, \alpha_2\}\myN - (\alpha_1+\alpha_2)\sqrt{\myN}
\end{split}
\label{transDiag}
\end{equation}
If the quadrant is processed so that we firstly perform the accumulations along the paths in \textit{p1)} starting at \(\myto=0\), and secondly we perform the accumulations along the paths in  \textit{p2)}, then:
\begin{itemize}
\item the first projection value along the paths in \ltextit{p1)} reaches the FC at \(\mytf{1}= \myJtilde\myTp\);
\item the last projection value along the paths in \ltextit{p1)} reaches the FC at most at \(\mytf{\myItilde+1}= (\myItilde\mytaui+\max\{\myItilde, \myJtilde\})\myTp\);
\item the first projection value along the paths in \ltextit{p2)} reaches the FC at most at \(\mytf{\myItilde+2}= \left[\left(\myItilde+1\right)\mytaui+\max\{\myItilde, \myJtilde\}\right]\myTp\);
\item the last projection value along the paths in \ltextit{p2)} reaches the FC at most at \(\mytf{\myItilde+\myJtilde+1}= \left[\left(\myItilde+\myJtilde+1\right)\mytaui+\myItilde\right]\myTp\).
\end{itemize}
Overall, the \(\pi/4\) oriented diagonal projections in a quadrant are performed after
\(
\left[\left(\myItilde+\myJtilde+1\right)\mytaui+\max\{\myItilde, \myJtilde\}\right]
\)
time slots.
Again, if the quadrants are processed in a serial fashion, the overall number of time slots needed to evaluate the diagonal projections sums up to
\begin{equation}
\begin{split}
&N^{(\pi/4)}_{\text{TS}}=2\left[(\myI+\myJ)\mytaui + (\max\{\myI,\myJ\}-1)\right]\\
&\phantom{N^{(\pi/4)}_{\text{TS}}}=2\sqrt{N}\left[(\alpha_1 + \alpha_2) \mytaui + \max\{\alpha_1, \alpha_2\}\right] -2
\end{split}
\label{slotsDiag}
\end{equation}
Fig.\ref{fig:diagtime} reports, for the sake of clarity, a time diagram summarizing the timing of projections evaluation when performing diagonal projections within a quadrant of the network.

Then, if we consider a Radon Like scheme comprising \(\myP=3\) projections along the directions \(\vartheta=0, \vartheta=\pi/2, \vartheta=\pi/4\), the overall number of needed transmissions sums up to (cfr. \eqref{transHor}, \eqref{transVer} and \eqref{transDiag}):
\begin{equation*}
\begin{split}
&N^{(0, \pi/2, \pi/4)}_{\text{TX}}\approx\gamma_3\myN\\
&\gamma_3=\left(2\alpha_1\alpha_2 + \dfrac{\alpha_1^2 + \alpha_2^2}{2}\right)+(\alpha_1+\alpha_2)\max\{\alpha_1, \alpha_2\}
\end{split}
\end{equation*}
while the number of needed time slots is (cfr. \eqref{slotsHor}, \eqref{slotsVer} and \eqref{slotsDiag}):
\begin{equation*}
\begin{split}
&N^{(0, \pi/2, \pi/4)}_{\text{TS}}\approx\delta_3\sqrt{\myN}\\
&\delta_3=\left(\alpha_1+\alpha_2\right)\left(4+4\mytaui\right) + \max\{\alpha_1, \alpha_2\}
\end{split}
\end{equation*}



To recap, the Radon Like CS  data gathering procedure let the fusion center collect all the needed measurement in a highly parallelized fashion; far from being optimal, the data  gathering scheme herein introduced allows a significant reduction in both the occupied bandwidth and the consumed energy \wrt state of the art data gathering scheme as the RS introduced in \cite{Fazel11}.
Further developments of globally optimized  Radon-like CS data gathering algorithms are still under investigation.

\end{document}